# Towards personalised assessment of abdominal aortic aneurysm structural integrity

Mostafa Jamshidian[1], Adam Wittek[1], Saeideh Sekhavat[1], Hozan Mufty[2, 3], Geert Maleux[4], Inge Fourneau[2, 3], Elke R. Gizewski[5], Eva Gassner[5], Alexander Loizides[5], Maximilian Lutz[5], Florian K. Enzmann[6], Donatien Le Liepvre[7], Florian Bernard[7], Ludovic Minvielle[7], Antoine Fondanèche[7], Karol Miller[1,*]

[1] Intelligent Systems for Medicine Laboratory, The University of Western Australia, Perth, Western Australia, Australia

[2] Department of Vascular Surgery, University Hospitals Leuven, Leuven, Belgium

[3] Department of Cardiovascular Sciences, Research Unit of Vascular Surgery, KU Leuven, Leuven, Belgium

[4] Department of Radiology, University Hospitals Leuven, Leuven, Belgium

[5] Department of Radiology, Medical University of Innsbruck, Innsbruck, Austria

[6] Department of Vascular Surgery, Medical University of Innsbruck, Innsbruck, Austria

[7] Nurea, Bordeaux, France

* Corresponding author.

E-mail address: karol.miller@uwa.edu.au (K. Miller).

## Abstract

Abdominal aortic aneurysm (AAA) is a life-threatening condition characterized by the progressive dilation of the aorta, which can lead to rupture if undetected or untreated. Stress-based rupture risk estimation using computational biomechanics has been widely studied; however, it requires wall strength data that cannot be measured in humans in vivo. To overcome this limitation, the goal of this study is to present a new method for biomechanical assessment of AAA via simultaneous consideration of tension and strain in AAA wall. We present a patient-specific, non-invasive method for assessing the structural integrity of the AAA wall using only time-resolved 3D computed tomography angiography (4D-CTA) images and blood pressure data. The proposed approach integrates wall strain (throughout the cardiac cycle) and wall tension analysis to compute a novel index, the Relative Structural Integrity Index (RSII), which quantifies local wall stiffness independently of wall thickness, wall material properties, and blood pressure measurement conditions. We applied our method to twenty patients from three different hospitals to extract visual RSII maps over the AAA wall of each individual patient and to compare the RSII values between aneurysmal and non-aneurysmal aortas in one patient. Our results primarily show similar RSII values across all patients, indicating the consistency of the method. Additionally, we observed patterns consistent with experimental findings reported in the literature: AAA walls exhibited higher stiffness than healthy aortic walls, while localized low-stiffness zones in the AAA wall were predominantly found in the most dilated regions.

## 1. Introduction

An abdominal aortic aneurysm (AAA) is a lasting and irreversible enlargement of the aorta that is typically asymptomatic, with diagnosis often occurring through incidental detection during imaging for some other condition. An untreated AAA can progressively enlarge and eventually rupture, which is fatal in most cases (NICE, 2020; Wanhainen et al., 2024).

The current approach to AAA management, which is based on the maximum diameter and growth rate, follows a one-size-fits-all approach that suggests clinical intervention when the aneurysm diameter exceeds 5.5 cm in men and 5 cm in women, or when the growth rate surpasses 1 cm per year (Wanhainen et al., 2024). The maximum diameter criterion may either underestimate or overestimate the rupture risk in individual AAA patients, as demonstrated by cases of ruptured AAAs with diameters below the critical diameter (Vorp, 2007) and stable, unruptured AAAs with diameters exceeding the critical diameter (Darling et al., 1977; NICE, 2020; Wanhainen et al., 2024). Autopsy results reveal that approximately 13% of AAAs with a maximum diameter of 5 cm or less ruptured, while 60% of those exceeding 5 cm may remain intact (Kontopodis et al., 2016). Ruptures of aneurysms smaller than 5 cm in patients under surveillance constitute avoidable disasters (most often patient dies) and are sometimes underreported.

AAA biomechanics, particularly wall stress analysis, has been widely studied as a potential tool to personalize disease management for individual patients (Chung et al., 2022; Farotto et al., 2018; Fillinger et al., 2003; Fillinger et al., 2002; Gasser et al., 2010; Gasser et al., 2022; Grassl et al., 2024; Indrakusuma et al., 2016; Joldes et al., 2016; Joldes et al., 2017; Li et al., 2008; Liddelow et al., 2023; Miller et al., 2020; Polzer et al., 2020; Singh et al., 2023; Speelman et al., 2007; Vande Geest et al., 2006; Wang et al., 2023). While biomechanical models have progressed to estimate patient-specific AAA wall tension (i.e., the stress resultant tangential to the aneurysm surface) without requiring information on the wall mechanical properties and thickness distribution (Joldes et al., 2016), stress- or tension-based rupture risk indicators rely on the assumption of the magnitude of the wall strength – the information not available for an individual patient, and usually derived from population-based data (Grassl et al., 2024; Singh et al., 2023). This is a major weakness of the current biomechanical approaches for personalized AAA rupture risk prediction and disease management, as direct non-invasive measurement of patient-specific in vivo wall strength is not feasible using currently available methods. However, wall stiffness or compliance can potentially be estimated by combining patient-specific wall stress (or tension) and strain data.

Several studies have investigated the non-invasive, in vivo measurement of AAA wall strain using sequential images captured at different phases of the cardiac cycle, commonly known as 4D imaging (Cebull et al., 2019; Derwich et al., 2023; Jamshidian et al., 2025b; Maas et al., 2024; Nagy et al., 2015; Raut et al., 2014; Satriano et al., 2015; Wang et al., 2018; Wittek et al., 2017; Wittek et al., 2018). Several studies (Nagy et al., 2015; Satriano et al., 2015) proposed that analysis of wall strain can be used to obtain patient-specific information about the aneurysm wall material properties and estimate the aneurysm wall rupture risk. Recently, we developed and validated a method for patient-specific, in vivo, and non-invasive AAA kinematic analysis by applying deformable image registration to time-resolved 3D computed tomography angiography (4D-CTA) images, enabling the measurement of wall strain throughout the cardiac cycle (Jamshidian et al., 2025b). To the best of our knowledge, no attempt has been made so far to perform an in vivo structural integrity analysis of the AAA wall by directly comparing stress (or tension) and strain.

The goal of this study is to present a new method for biomechanical assessment of AAA via simultaneous consideration of tension and strain in AAA wall. Despite decades of effort, AAA wall stress has not been demonstrated to be a reliable biomarker for patient stratification; therefore, new approaches are needed. We posit that simultaneous analysis of AAA wall stress and strain may lead to progress in the field. Accordingly, we propose a novel biomechanical metric, which integrates wall tension and wall strain, thereby transcending the conventional stress-based analyses reported in the literature.

In this study, we combine the AAA wall strain measurement with the patient-specific calculation of tension within the AAA wall to formulate a novel Relative Structural Integrity Index (RSII) as a potential indicator for assessing the severity of AAA disease. We calculate the distribution of this index on the AAA surface for twenty patients, whose 4D CTA images were sourced from Fiona Stanley Hospital in Australia, Medical University of Innsbruck in Austria, and University Hospitals Leuven in Belgium. RSII takes into account both the tension in the AAA wall, computed using an established finite element (FE)-based stress recovery technique (Joldes et al., 2016; Joldes et al., 2017; Joldes et al., 2018), and the strain derived from deformable image registration of individual 3D frames of 4D-CTA image, thereby integrating wall internal force information with AAA kinematics (Jamshidian et al., 2025b).

The paper is organized as follows: In Section 2, we present the image data of AAA patients and the methods for estimating wall stress, tension and strain, and then introduce our patient-specific RSII of the AAA wall. In Section 3, we present the results for AAA wall tension, strain, and RSII in twenty patients, followed by conclusions and discussion in Section 4.

## 2. Materials and methods

### 2.1. Image data

We used anonymized contrast-enhanced 4D-CTA image datasets from twenty AAA patients, with each dataset consisting of up to ten 3D volume frames per cardiac cycle. Patients 1 to 10 were recruited at Fiona Stanley Hospital (Perth, Australia), patients 11 to 17 at Medical University of Innsbruck (Innsbruck, Austria), and patients 18 to 20 at University Hospitals Leuven (Leuven, Belgium), with informed consent obtained before their participation. The study was conducted in accordance with the Declaration of Helsinki, and the protocols were approved by Human Research Ethics and Governance at South Metropolitan Health Service (HREC-SMHS) (approval code RGS3501), Human Research Ethics Office at The University of Western Australia (approval code RA/4/20/5913), Ethics Committee of Medical University of Innsbruck (approval code 1271/2023), and Ethics Committee of University Hospitals Leuven (approval code S63163).

Table 1 provides the image dimensions and resolutions for each patient, as well as the maximum AAA diameter, which is currently the primary criterion for surgical intervention. The image size and resolution varied among individuals depending on the permitted safe radiation dose, which was determined by the patient’s weight and height.

As an example of the image data, Figure 1a shows the cropped 3D-CTA image of Patient 1's AAA in systolic phase. The image dimensions are 72 x 61 x 83 voxels, with voxel spacing of 1.18 mm x 1.18 mm x 1.00 mm along the R (left-right direction), A (posterior-anterior direction), S (inferior-superior direction) axes, as illustrated in Figure 1. In the patient coordinate system, the basis vectors align with the anatomical axes: anterior-posterior, inferior-superior, and left-right.

**Table 1.** Computed tomography angiography (CTA) image dimensions and resolutions for twenty patients with abdominal aortic aneurysm (AAA), along with the maximum AAA diameter for each patient. Patients 1 to 10 were from Fiona Stanley Hospital (Perth, Australia), patients 11 to 17 from Medical University of Innsbruck (Innsbruck, Austria), and patients 18 to 20 from University Hospitals Leuven (Leuven, Belgium).

| Patient No. | Image dimensions (pixels) | Image resolution (mm) | AAA maximum diameter (mm) |
|---|---|---|---|
| 1 | 256 × 256 × 254 | 1.18 × 1.18 × 1.00 | 65.8 |
| 2 | 512 × 512 × 169 | 0.63 × 0.63 × 1.50 | 67.2 |
| 3 | 512 × 512 × 143 | 0.63 × 0.63 × 1.00 | 50.9 |
| 4 | 256 × 256 × 177 | 0.81 × 0.81 × 1.50 | 59.3 |
| 5 | 256 × 256 × 482 | 1.82 × 1.82 × 1.00 | 52.7 |
| 6 | 256 × 256 × 488 | 1.53 × 1.53 × 1.00 | 47.1 |
| 7 | 512 × 512 × 160 | 0.63 × 0.63 × 1.00 | 57.3 |
| 8 | 512 × 512 × 155 | 0.31 × 0.31 × 1.00 | 44.2 |
| 9 | 256 × 256 × 170 | 1.25 × 1.25 × 1.00 | 50.2 |
| 10 | 256 × 256 × 184 | 1.25 × 1.25 × 1.00 | 52.3 |
| 11 | 512 × 512 × 137 | 0.23 × 0.23 × 1.00 | 65.0 |
| 12 | 512 × 512 × 246 | 0.39 × 0.39 × 0.70 | 61.3 |
| 13 | 512 × 512 × 206 | 0.23 × 0.23 × 1.00 | 57.4 |
| 14 | 512 × 512 × 206 | 0.23 × 0.23 × 1.00 | 53.3 |
| 15 | 512 × 512 × 172 | 0.25 × 0.25 × 1.00 | 53.4 |
| 16 | 512 × 512 × 137 | 0.23 × 0.23 × 1.00 | 50.0 |
| 17 | 512 × 512 × 172 | 0.23 × 0.23 × 1.00 | 56.7 |
| 18 | 512 × 512 × 171 | 0.31 × 0.31 × 0.70 | 53.3 |
| 19 | 512 × 512 × 191 | 0.31 × 0.31 × 0.70 | 51.3 |
| 20 | 512 × 512 × 151 | 0.31 × 0.31 × 0.70 | 54.5 |

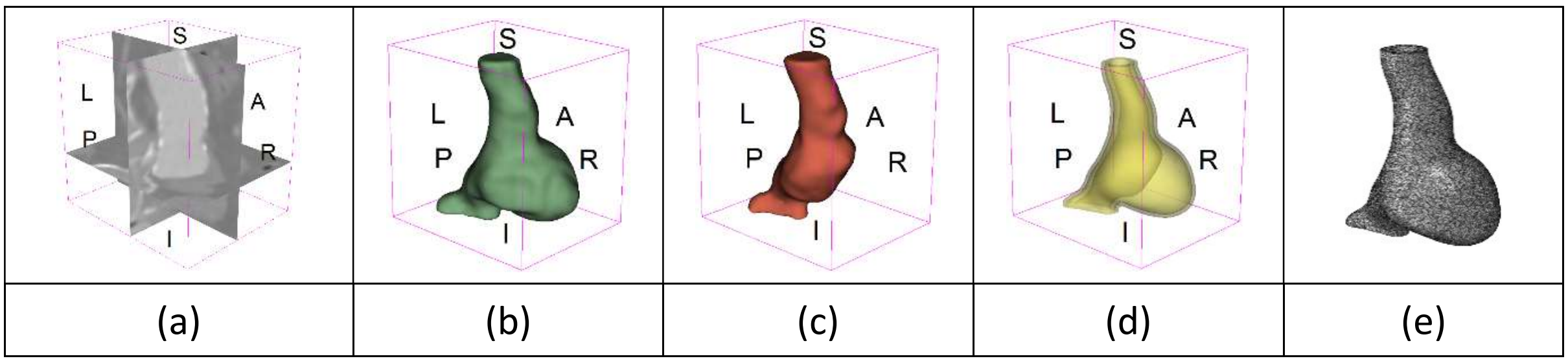

**Figure 1.** Patient 1's abdominal aortic aneurysm (AAA): (a) Cropped 3D-CTA image, (b) AAA label map, (c) Blood label map, (d) Surface model of AAA including ILT, and (e) Finite element (FE) mesh.

### 2.2. Stress and tension

For stress and tension computations, we used BioPARR (Biomechanics-based Prediction of Aneurysm Rupture Risk), a free software package for AAA biomechanical analysis based on the FE method (https://bioparr.mech.uwa.edu.au/) (Joldes et al., 2017). Given the patient's blood pressure, along with the AAA and blood segmentations (label maps) extracted from medical images, BioPARR performs automated, efficient, and robust wall stress recovery using the linear FE method and integrates the tangential components of the stress tensor over the wall thickness to obtain a stress resultant, i.e., the wall tension (N/m) that, unlike stress (Miller et al., 2018), is independent of very difficult to measure AAA wall thickness.

This efficient methodology enables rapid computation of the wall tension that balances the applied blood pressure load (in the deformed configuration), without requiring the mechanical properties of the wall tissue nor its thickness (Ciarlet, 2021; Joldes et al., 2016; Joldes et al., 2017). Alternatively, a sophisticated non-linear inverse methodology can be applied to recover stress and tension by "unloading" an AAA (see, e.g., (Joldes et al., 2015; Lu et al., 2007, 2008; Miller and Lu, 2013)), but our experience shows that these complicated methods yield the results that are, for practical purposes, the same as those obtained using our simple stress recovery approach (Joldes et al., 2016), as predicted by the theory (Ciarlet, 2021).

To extract AAA geometry for patients 1 to 10, we used PRAEVAorta by NUREA (https://www.nurea-soft.com/) for AI-based automatic segmentation of 3D-CTA image (see Figure 1a), followed by automated post-processing with an in-house MATLAB code (Alkhatib et al., 2024), to automatically generate the AAA and blood label maps, as shown in Figure 1b and Figure 1c, respectively. The 4D-CTA images and AAA models for patients 1 to 10 are publicly available in our Zenodo database "4D-CTA Images and Surface Models of Abdominal Aortic Aneurysms from Ten Patients" (Jamshidian et al., 2025a).

To extract AAA geometry for patients 11 to 20, we used the nnInteractive extension (de Vente et al., 2025) in the 3D Slicer image computing platform (Fedorov et al., 2012) for AI-assisted segmentation of 3D-CTA image, to generate the AAA and blood label maps. The nnInteractive is a 3D promptable and open-set AI-based segmentation method that enables users to provide intuitive inputs such as points and scribbles, which are then transformed into high-quality volumetric segmentations. Unlike traditional slice-by-slice approaches, nnInteractive directly generates 3D label maps from sparse 2D prompts.

Using the AAA and blood label maps and the assumed wall thickness of 1.5 mm (Raut et al., 2013), we then ran BioPARR to automatically extract the wall and intraluminal thrombus (ILT) geometries, create the AAA surface model (Figure 1d), generate FE mesh, fix the top and bottom ends of the AAA, apply a uniform pressure of 13 kPa to the inner surface of the ILT, call Abaqus/Standard FE package (Simulia, 2024) to recover stress using the FE method, incorporate residual stresses using the Fung's Uniform Stress Hypothesis (Fung, 1991; Joldes et al., 2018) and finally integrate tangential components of stress (therefore "integrating out" the assumed wall thickness) to output the wall tension.

The mesh generated using BioPARR, shown in Figure 1e, consists of 824579 nodes and 481928 10-node quadratic tetrahedron elements with hybrid formulation and constant pressure (element type C3D10H in Abaqus) (Alkhatib et al., 2023). For stress recovery, BioPARR assumes isotropic linear elastic material properties for the wall tissue, with a Young's modulus of 100 GPa and a Poisson's ratio of 0.49. The ILT was assumed to be 20 times more compliant than the wall, an estimate previously used by Miller et al. (2020) based on data from (Di Martino et al., 2006; Kavanagh et al., 2014; Tong et al., 2011). This assumption is somewhat arbitrary, but the results are only weakly sensitive to the selected value of thrombus stiffness, provided it remains much lower than that of the AAA wall. The high stiffness preserves the shape of the observed aneurysm, ensuring that the stress distribution corresponds to the observed, loaded structure. This approach avoids a common mistake in the literature—applying pressure load to an already deformed geometry, de facto treating it as unloaded. This results in computing stress distribution in an unphysiological, excessively inflated AAA geometry.

BioPARR incorporates residual stresses based on Fung's Uniform Stress Hypothesis (Fung, 1991; Joldes et al., 2018), making it an easy post-processing step in AAA stress analysis (Joldes et al., 2018). We previously confirmed that, for a given material model and properties, the residual stress fields generated by the BioPARR method closely match those obtained through the more complex non-linear iterative stress analysis methods (Joldes et al., 2018; Polzer et al., 2013). Finally, in another post-processing step, wall tension is computed by integrating the tangential components of the stress tensor over the wall thickness, resulting in a stress resultant independent of the assumed wall thickness.

**2.3. Strain**

To compute AAA wall displacement and strain, we applied deformable image registration to align the systolic and diastolic 3D frames of the 4D-CTA, estimating the displacement field that maps the systolic to the diastolic AAA geometry. Next, we derived wall displacements from the registration displacement field and then calculated wall strain.

We used a MATLAB implementation of deformable image registration with isotropic total variation regularization of displacement (Jamshidian et al., 2025b; Vishnevskiy et al., 2017). For details on the image registration theory and algorithm, the reader may refer to our recent work on AAA kinematics (Jamshidian et al., 2025b).

We interpolated the registration displacement field at the AAA wall, represented by a point cloud, to obtain the wall displacement from registration in the Cartesian patient coordinate system (R, A, S). We have previously shown that registration produces more accurate displacements in the direction of the image gradient, which, in the case of the wall region of AAA 3D-CTA images, closely aligns with the wall normal (Jamshidian et al., 2025b; Lehoucq et al., 2021). Thus, we established a local biological coordinate system consisting of the local normal and two perpendicular tangents to the wall surface using 3D plane fitting via the planar least squares regression method (Berkmann and Caelli, 1994; Shakarji, 1998) and decomposed the wall displacement into its normal $u_n$ and tangential $u_t$ components.

Wittek et al. (2018) used the FE method to compute the AAA wall strain field from 4D ultrasound wall motion data and demonstrated that circumferential strain is the predominant and physiologically meaningful measure of AAA wall strain. Similarly, mechanical testing of uniaxial samples harvested during open AAA repair along axial and circumferential directions revealed that wall strength in the circumferential direction is significantly higher than in the axial direction, further confirming the significance and dominance of circumferential strain (Polzer et al., 2021).

Following the methods of (Jamshidian et al., 2025b; Karatolios et al., 2013; Morrison et al., 2009), we used the wall normal displacement from registration, $u_n$, to calculate the local circumferential wall strain as (Coelho et al., 2014; Fichter, 1997; Hencky, 1915):

$$\epsilon = \frac{u_n}{R} \tag{1}$$

where $R$ is the local radius of wall curvature, estimated by local surface fitting with non-deterministic outlier detection (Jamshidian et al., 2025b; Torr and Zisserman, 2000).

### 2.4. Structural Integrity Index

Attarian et al. (2019) observed longitudinal rupture lines in fully harvested ruptured AAAs and used crack propagation simulations to show that wall stress alone was unlikely to cause rupture. They concluded that rupture likely began in short segments of less than 1 cm and then propagated along the observed rupture lines.

The longitudinal rupture lines indicate the dominance of circumferential strain and wall tension in AAA rupture mechanism. Therefore, we defined a Structural Integrity Index (SII) for the AAA wall as the ratio of circumferential strain to wall tension $t$ (N/m), i.e.,

$$\mathrm{SII} = \frac{\epsilon}{t} \tag{2}$$

with units of m/N. SII serves as an indicator of the local wall compliance and offers a visual map of its variations. By quantifying the localized weakening of the wall, SII visualizes regions at risk of rupture as regions of the AAA wall with relatively high SII values compared to the surrounding regions.

As a new integrated measure of AAA wall compliance, we defined and computed the Relative Structural Integrity Index (RSII) as the SII value at a given point divided by the average absolute value of SII across the AAA wall surface, i.e.,

$$\mathrm{RSII} = \frac{|\mathrm{SII}|}{\mathrm{mean}(|\mathrm{SII}|)} \tag{3}$$

which may serve as a novel indicator of AAA structural soundness, the severity of AAA disease, and possibly a predictor of AAA disease progression, including rupture risk. As a relative measure, RSII is independent of blood pressure measurement accuracy. Further details on SII are provided in the Discussion section.

## 3. Results

### 3.1. AAA structural integrity

We applied our methods to evaluate AAA stress, tension, strain, and RSII in twenty patients. For each patient, we used the diastolic and systolic phases of 4D-CTA image data as the moving and fixed images in deformable image registration, respectively. For image registration parameters, see our recent work on AAA kinematics (Jamshidian et al., 2025b).

Table 2 summarizes the AAA structural integrity analysis results for twenty patients. Table 2 presents the AAA geometry and contour plots of wall tension, circumferential strain, SII and RSII for each patient, using patient-specific contour limits.

**Table 2.** AAA structural integrity analysis results for twenty patients. For each patient, the AAA geometry including ILT, and contour plots of wall tension, circumferential strain, Structural Integrity Index (SII) and Relative Structural Integrity Index (RSII) are shown. Patient-specific contour limits, including the 99th percentile wall tension $t_{\circ}$, the 99th percentile circumferential strain $\epsilon_{\circ}$, the 99th percentile structural integrity index $SII_{\circ}$ and the 99th percentile relative structural integrity index $RSII_{\circ}$, are reported for each patient.

| Patient number | AAA geometry including ILT | Wall tension (N/mm)<br>$0$ … $+t_{\circ}$<br>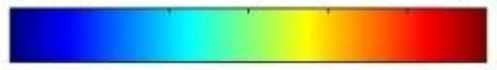 | Circumferential strain (%)<br>$-\epsilon_{\circ}$ … $0$ … $+\epsilon_{\circ}$<br>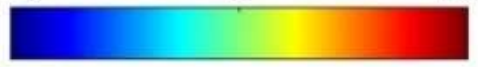 | SII (mm/N)<br>$-\mathrm{SII}_{\circ}$ … $0$ … $+\mathrm{SII}_{\circ}$<br>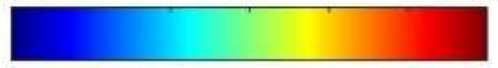 | RSII<br>$0$ … $+\mathrm{RSII}_{\circ}$<br>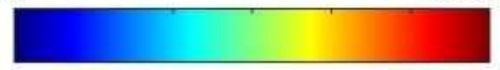 |
|---|---|---|---|---|---|
| 1 | 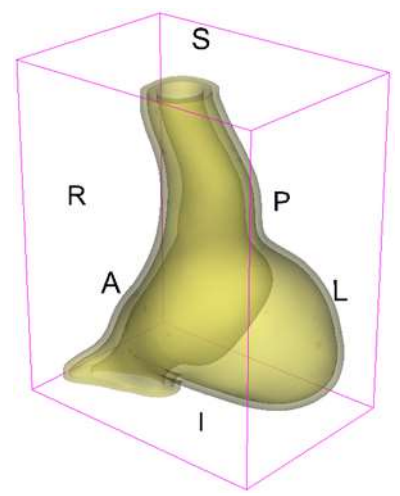 | 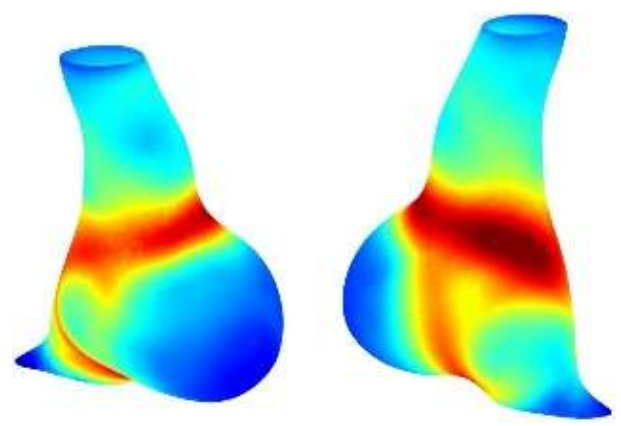<br>$t_{\circ} = 0.27\ \mathrm{N/mm}$ | 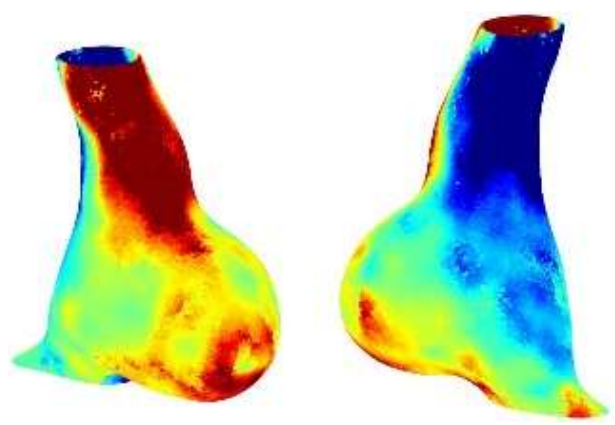<br>$\epsilon_{\circ} = 4.78\ \%$ | 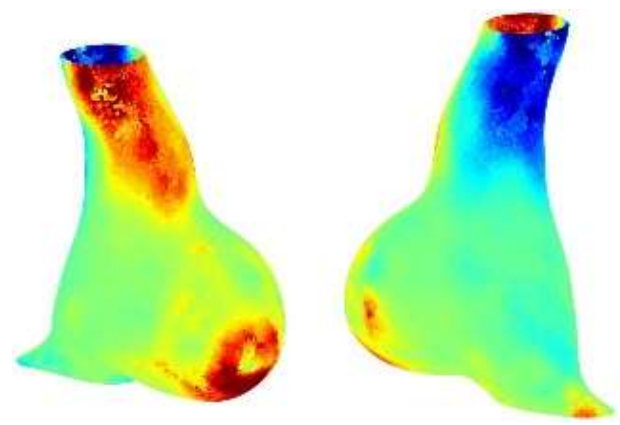<br>$\mathrm{SII}_{\circ} = 1.21\ \mathrm{mm/N}$ | 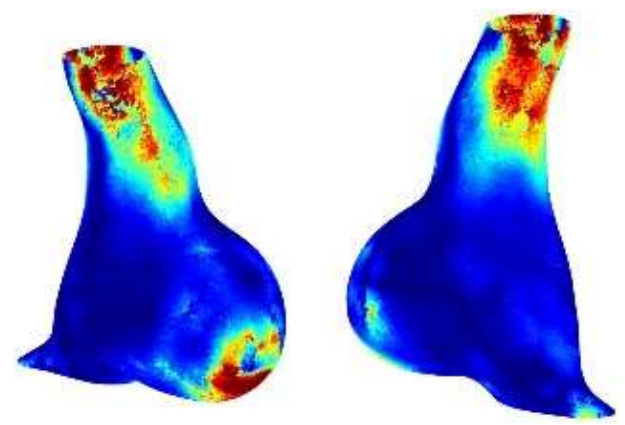<br>$\mathrm{RSII}_{\circ} = 4.85$ |
| 2 | 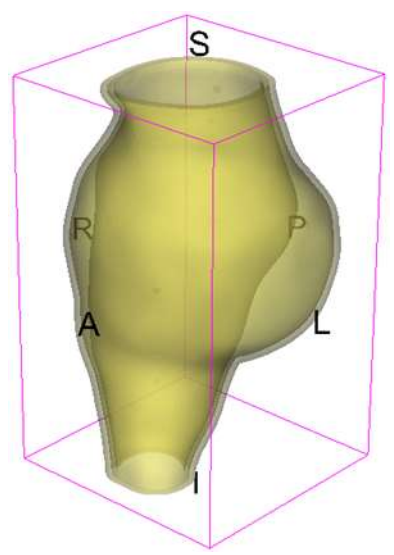 | 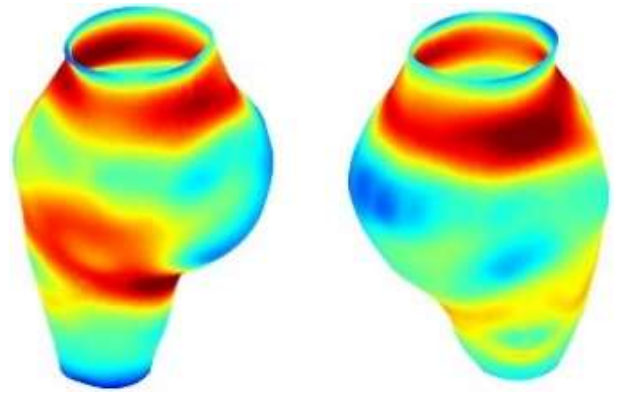<br>$t_{\circ} = 0.44\ \mathrm{N/mm}$ | 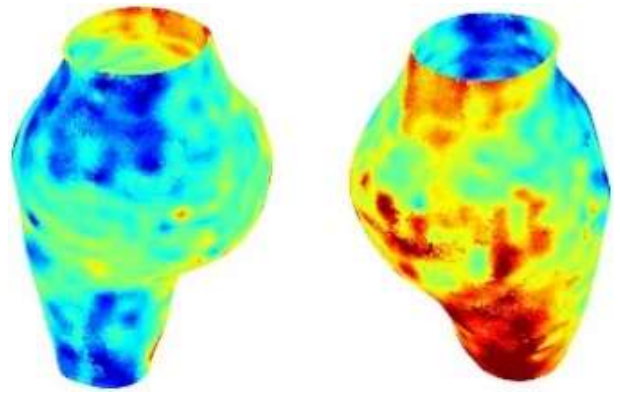<br>$\epsilon_{\circ} = 5.29\ \%$ | 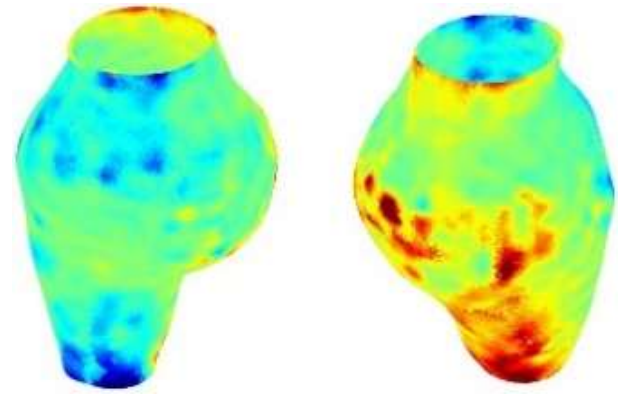<br>$\mathrm{SII}_{\circ} = 0.37\ \mathrm{mm/N}$ | 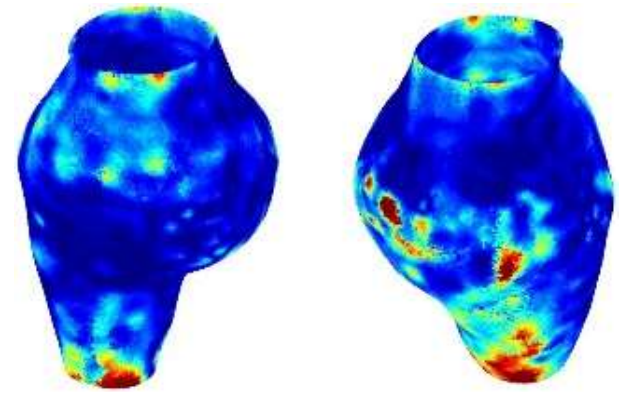<br>$\mathrm{RSII}_{\circ} = 4.83$ |
| 3 | 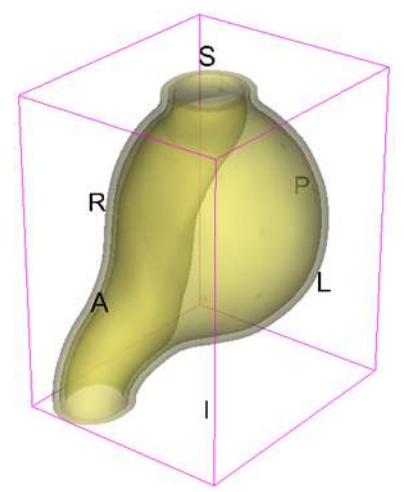 | 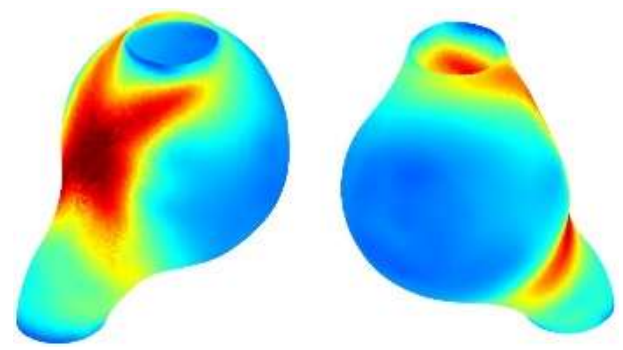<br>$t_\circ = 0.29$ N/mm | 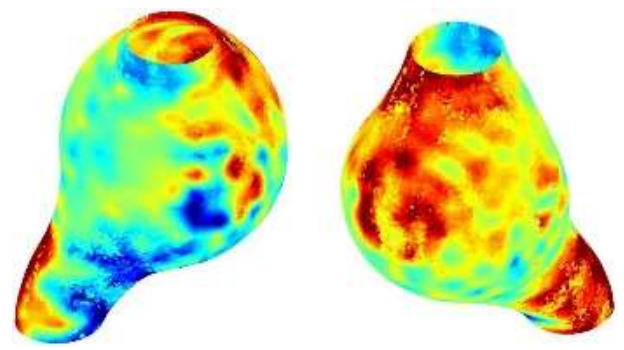<br>$\epsilon_\circ = 5.26$ % | 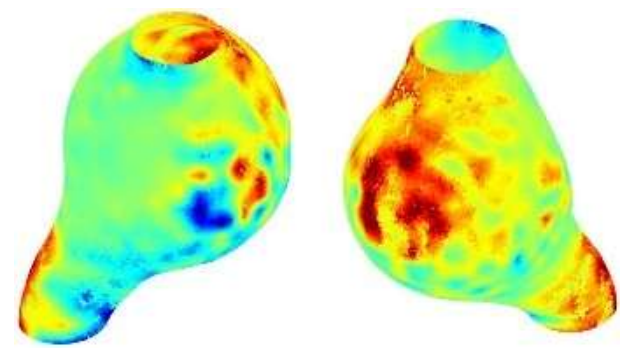<br>$\mathrm{SII}_\circ = 0.75$ mm/N | 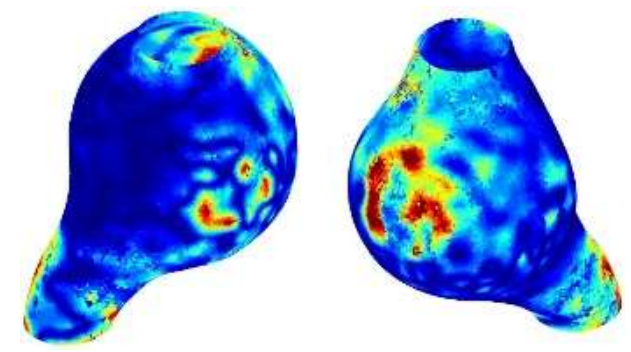<br>$\mathrm{RSII}_\circ = 4.80$ |
| 4 | 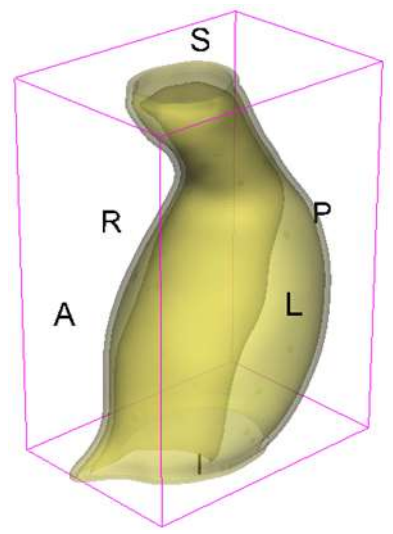 | 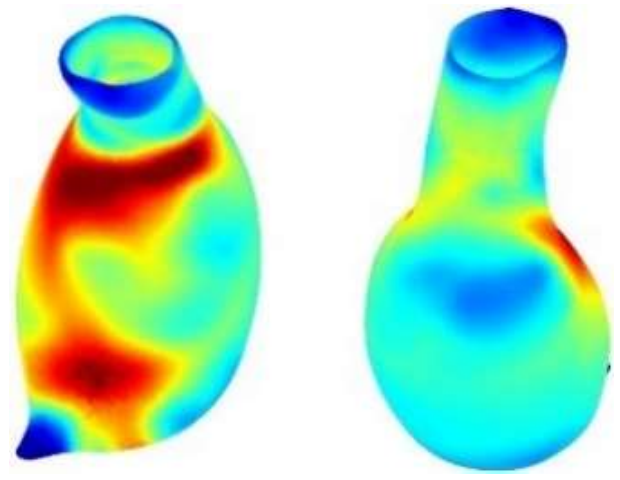<br>$t_\circ = 0.38$ N/mm | 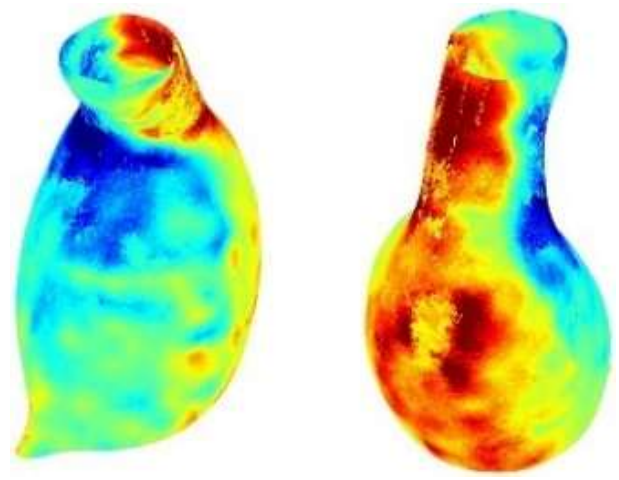<br>$\epsilon_\circ = 4.77$ % | 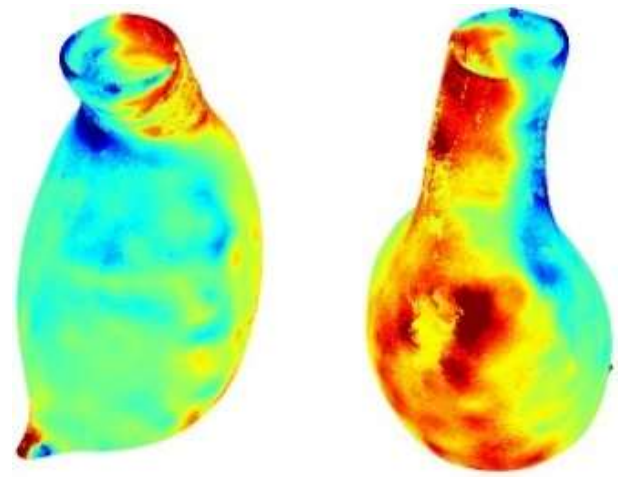<br>$\mathrm{SII}_\circ = 0.36$ mm/N | 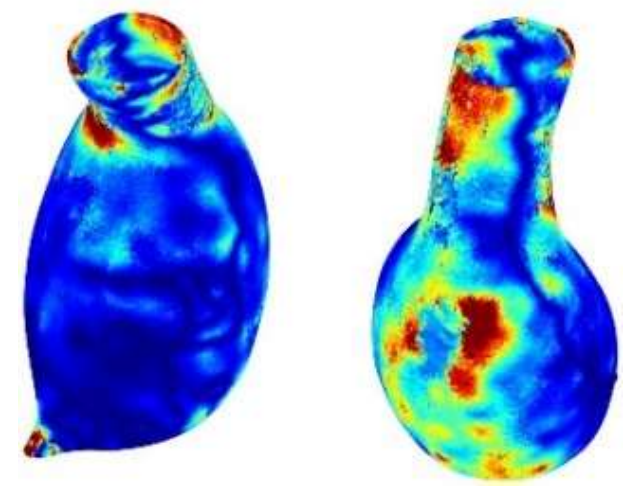<br>$\mathrm{RSII}_\circ = 3.57$ |
| 5 | 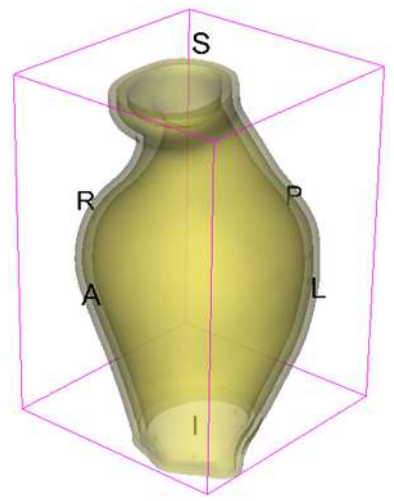 | 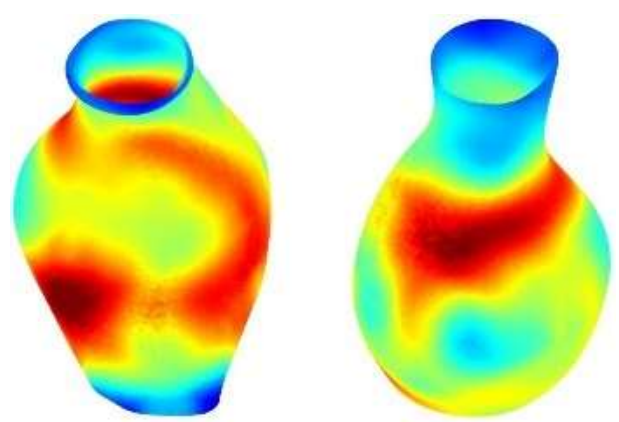<br>$t_\circ = 0.38$ N/mm | 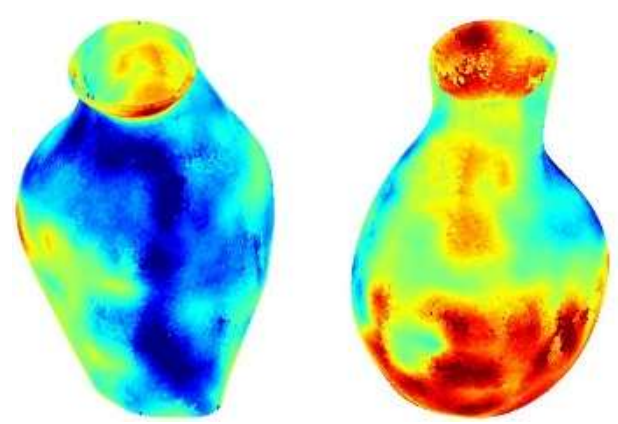<br>$\epsilon_\circ = 4.22$ % | 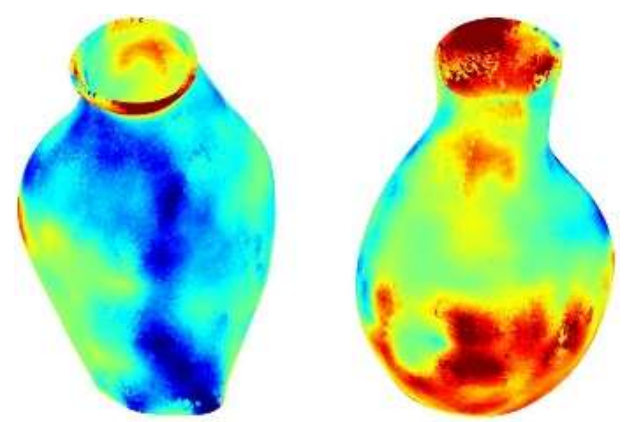<br>$\mathrm{SII}_\circ = 0.24$ mm/N | 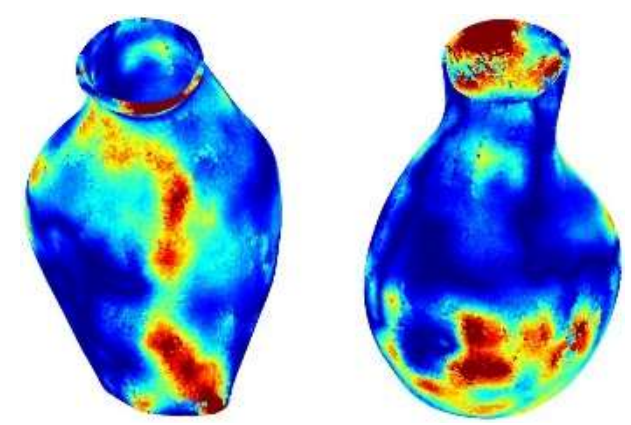<br>$\mathrm{RSII}_\circ = 3.22$ |
| 6 | 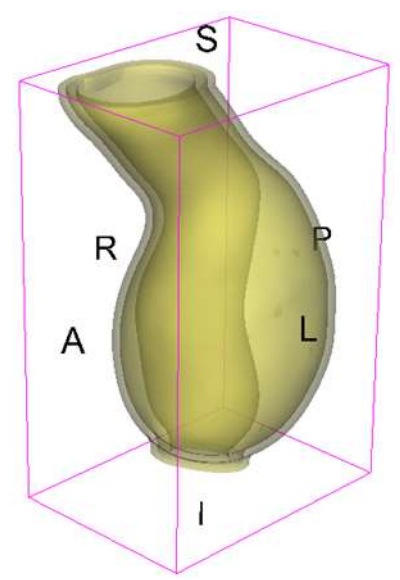  | 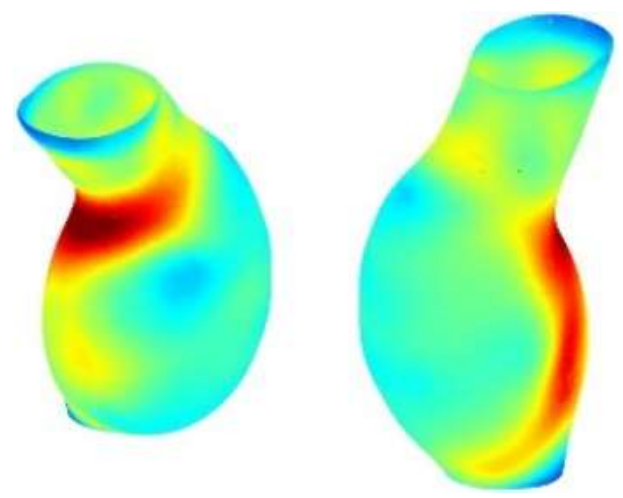 $t_\circ = 0.29$ N/mm | 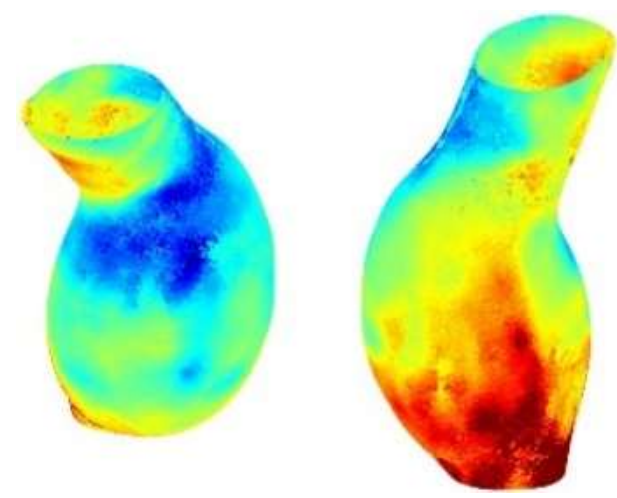 $\epsilon_\circ = 4.98$ % | 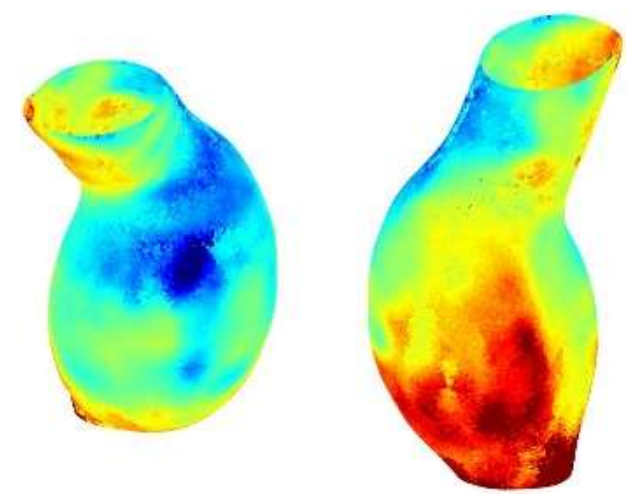 $\mathrm{SII}_\circ = 0.32$ mm/N | 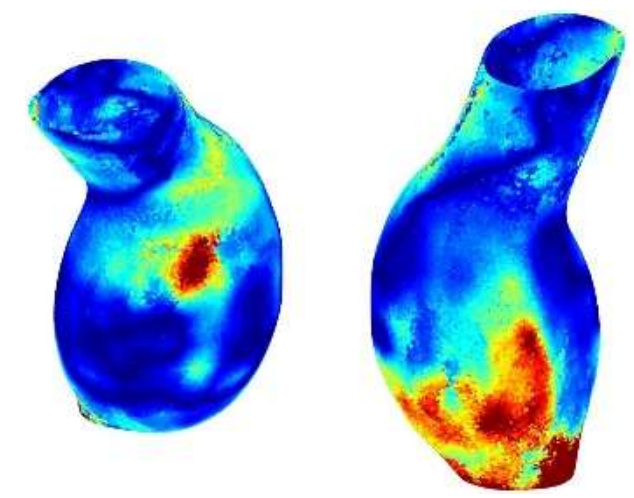 $\mathrm{RSII}_\circ = 3.22$ |
| 7 | 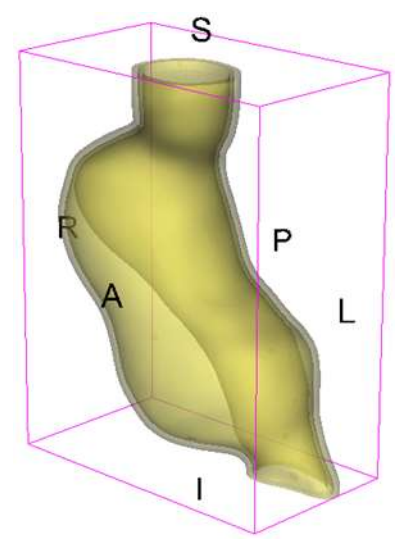  | 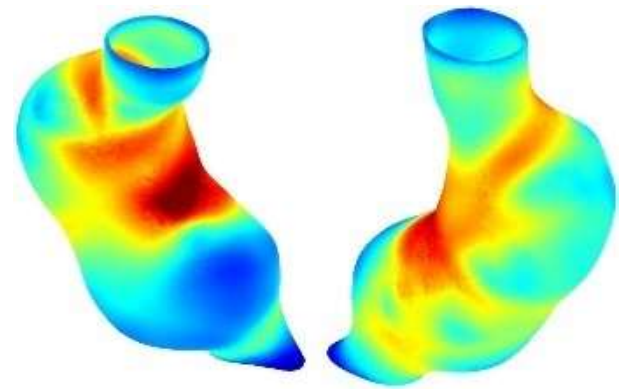 $t_\circ = 0.38$ N/mm | 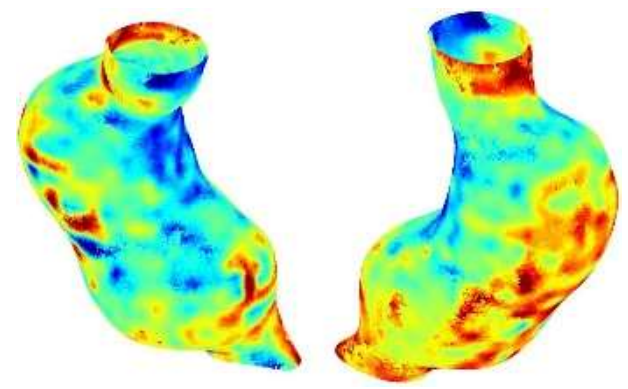 $\epsilon_\circ = 5.94$ % | 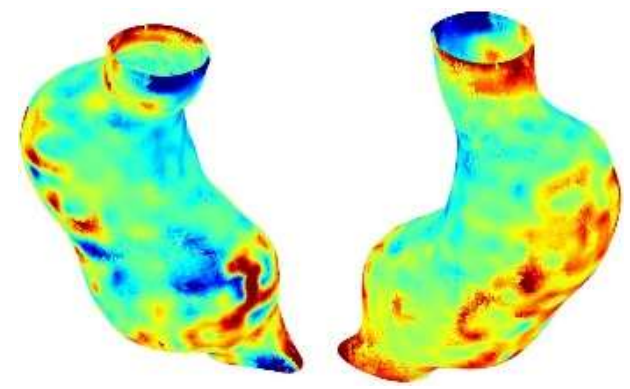 $\mathrm{SII}_\circ = 0.39$ mm/N | 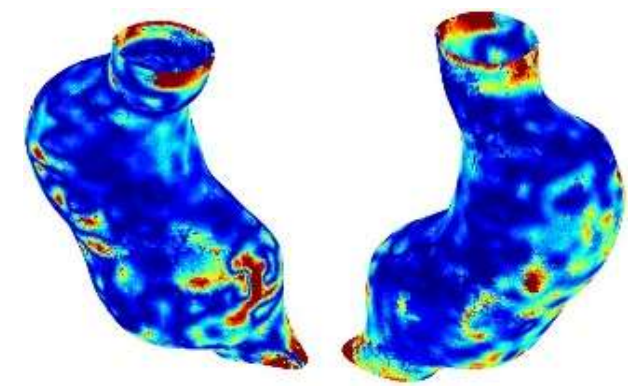 $\mathrm{RSII}_\circ = 3.66$ |
| 8 | 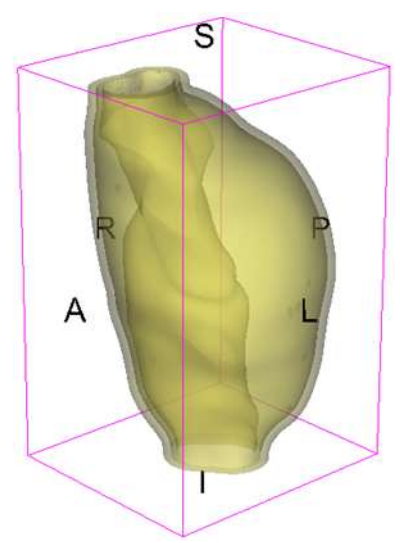  | 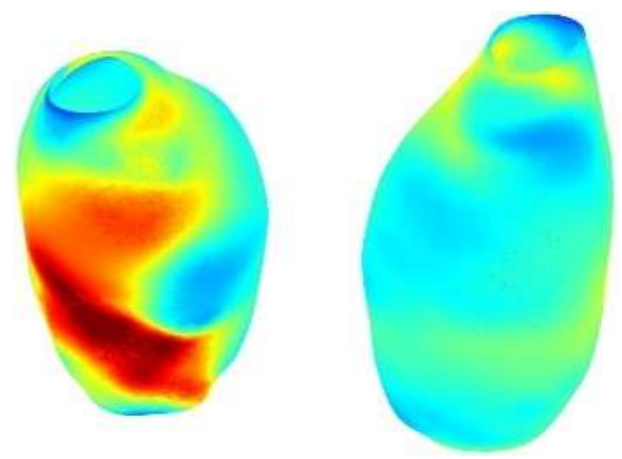 $t_\circ = 0.28$ N/mm | 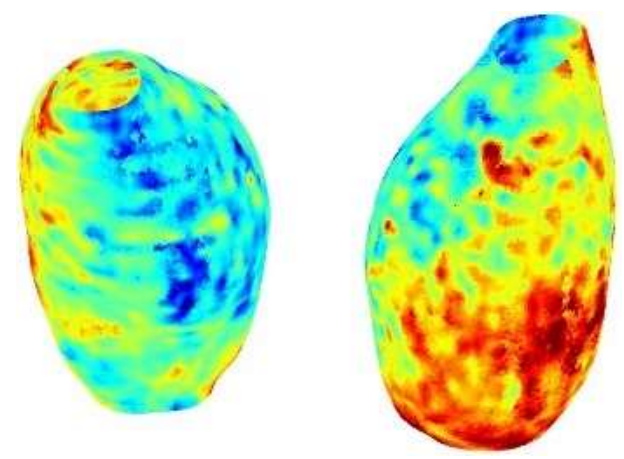 $\epsilon_\circ = 4.56$ % | 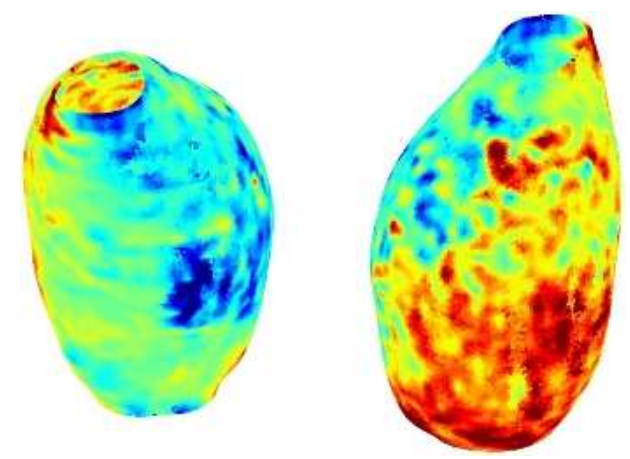 $\mathrm{SII}_\circ = 0.35$ mm/N | 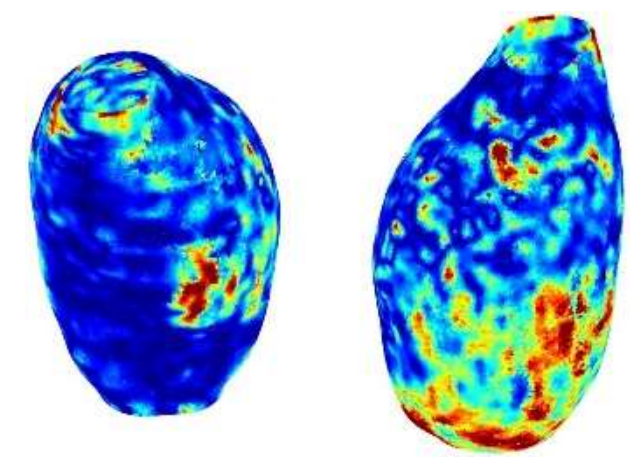 $\mathrm{RSII}_\circ = 3.32$ |
| 9 | 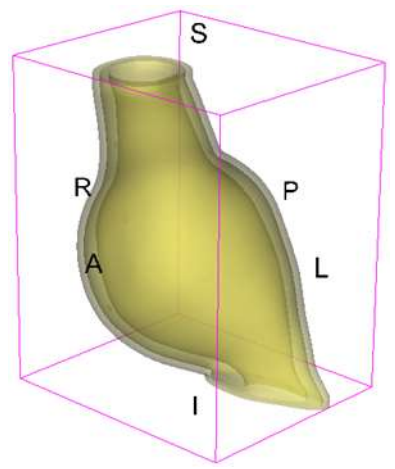 | 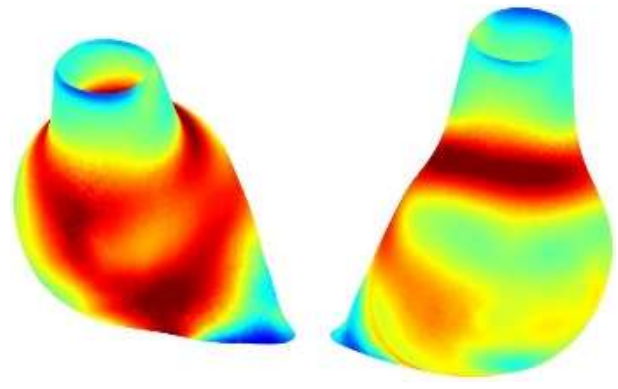 $t_{\circ} = 0.29$ N/mm | 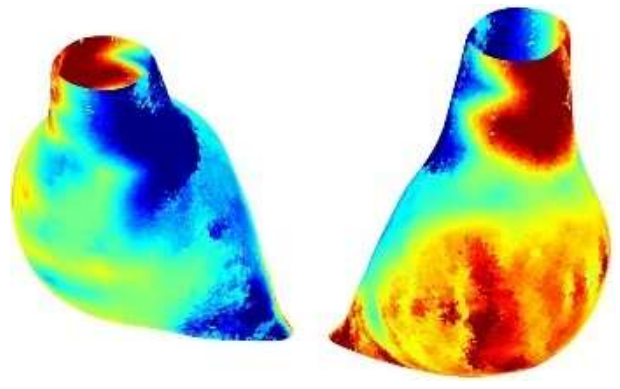 $\epsilon_{\circ} = 2.88$ % | 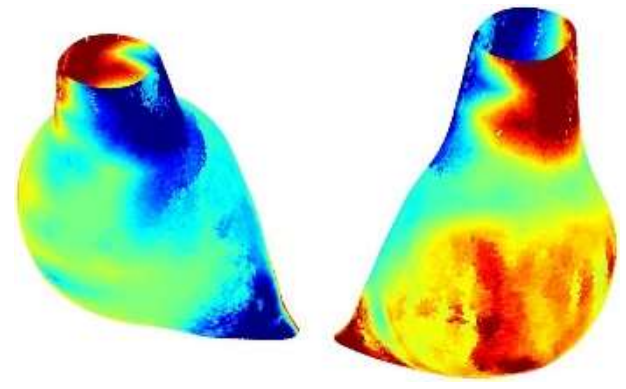 $\mathrm{SII}_{\circ} = 0.20$ mm/N | 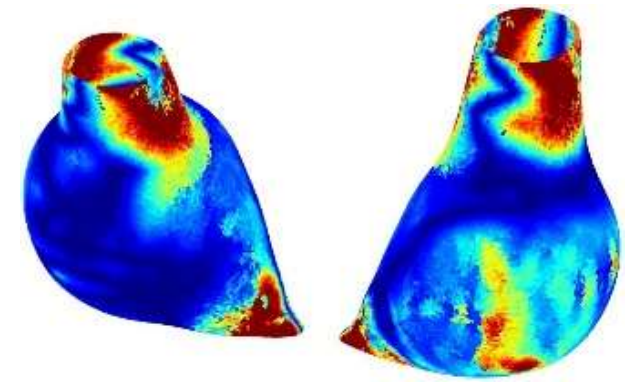 $\mathrm{RSII}_{\circ} = 3.08$ |
| 10 | 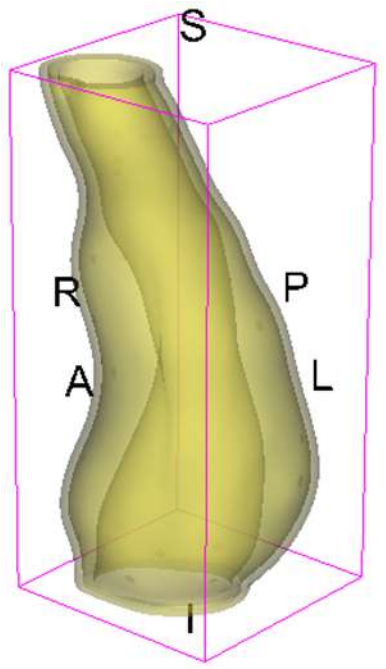 | 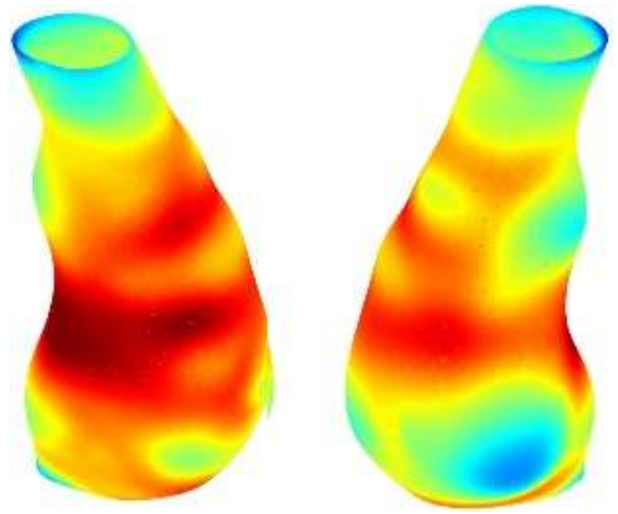 $t_{\circ} = 0.25$ N/mm | 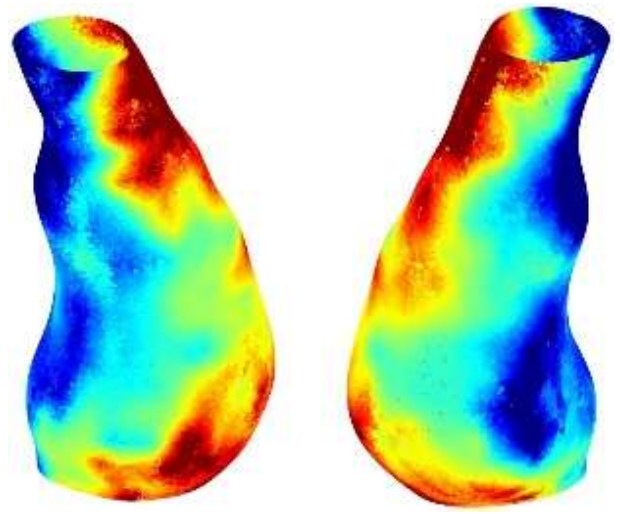 $\epsilon_{\circ} = 4.29$ % | 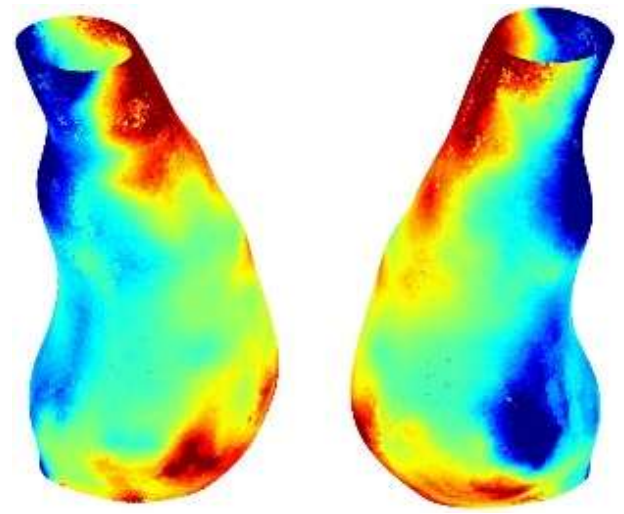 $\mathrm{SII}_{\circ} = 0.35$ mm/N | 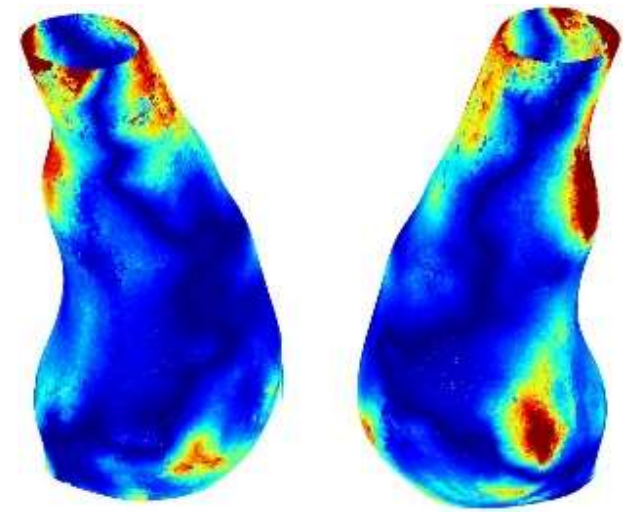 $\mathrm{RSII}_{\circ} = 3.47$ |
| 11 | 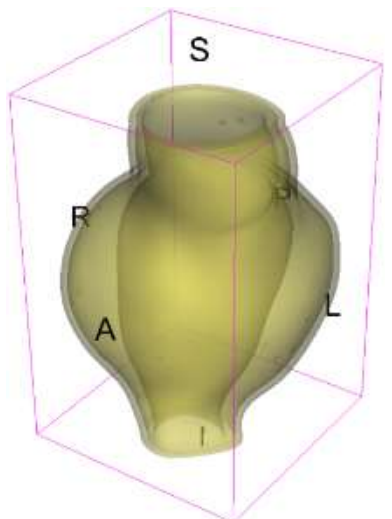 | 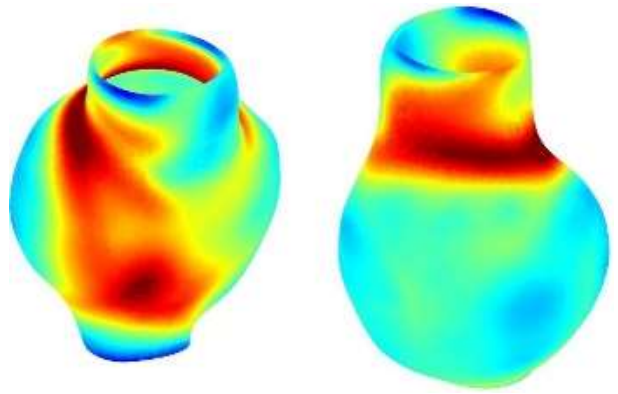 $t_{\circ} = 0.37$ N/mm | 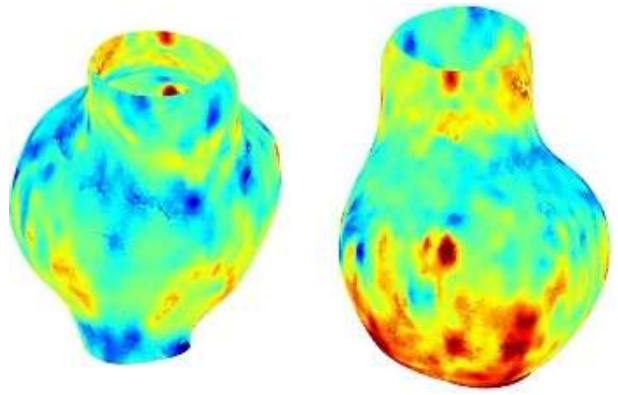 $\epsilon_{\circ} = 7.75$ % | 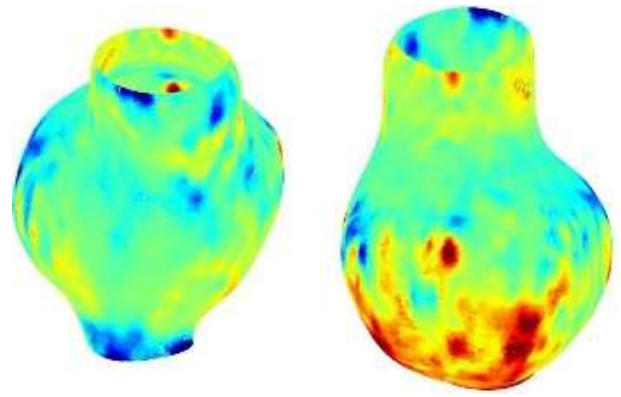 $\mathrm{SII}_{\circ} = 0.56$ mm/N | 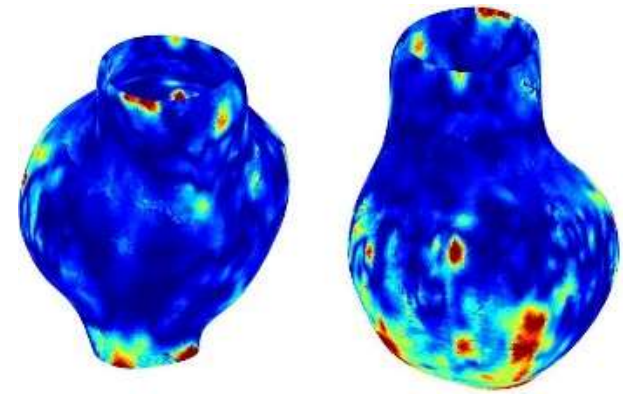 $\mathrm{RSII}_{\circ} = 4.82$ |
| 12 | 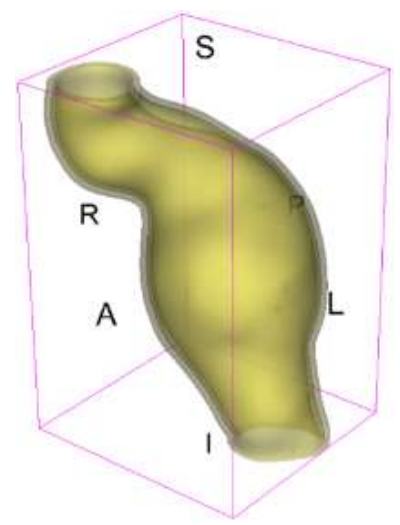 | 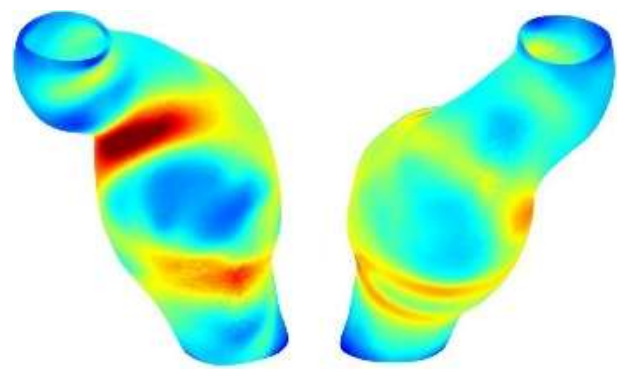<br>$t_\circ = 0.49\ \mathrm{N/mm}$ | 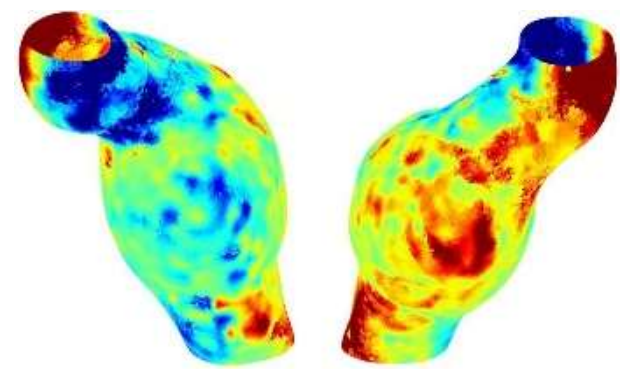<br>$\epsilon_\circ = 5.14\ \%$ | 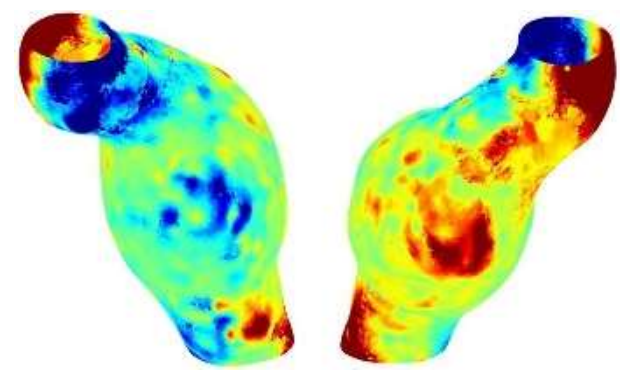<br>$\mathrm{SII}_\circ = 0.28\ \mathrm{mm/N}$ | 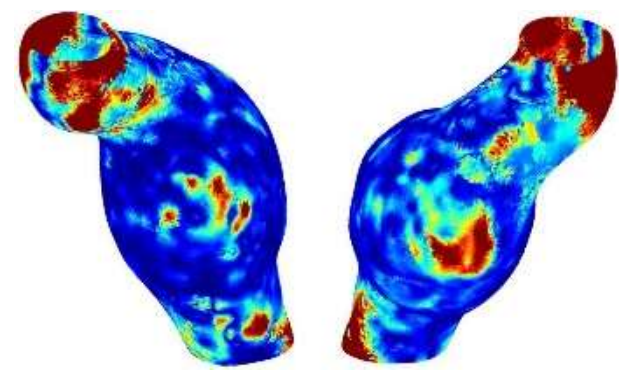<br>$\mathrm{RSII}_\circ = 3.08$ |
| 13 | 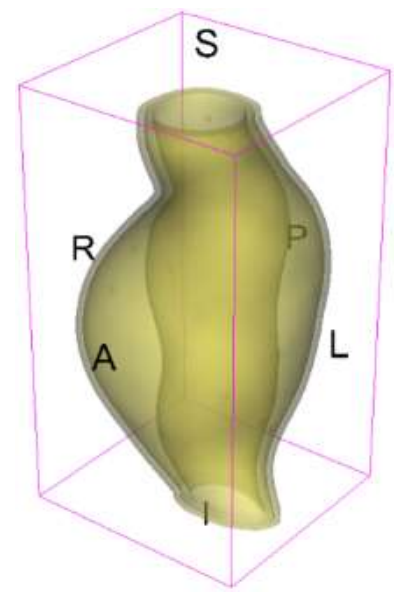 | 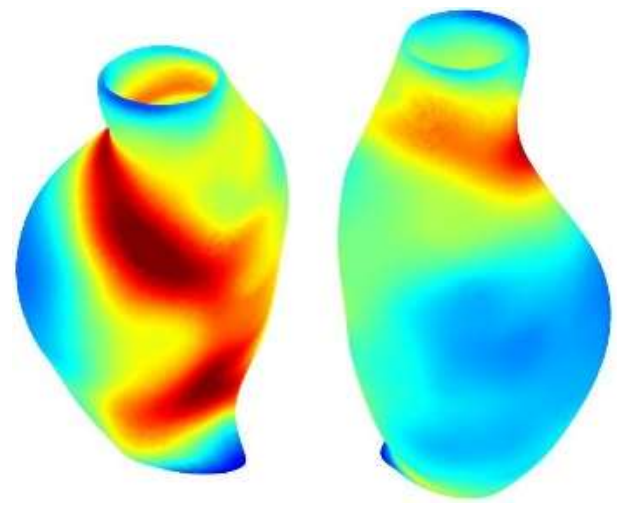<br>$t_\circ = 0.32\ \mathrm{N/mm}$ | 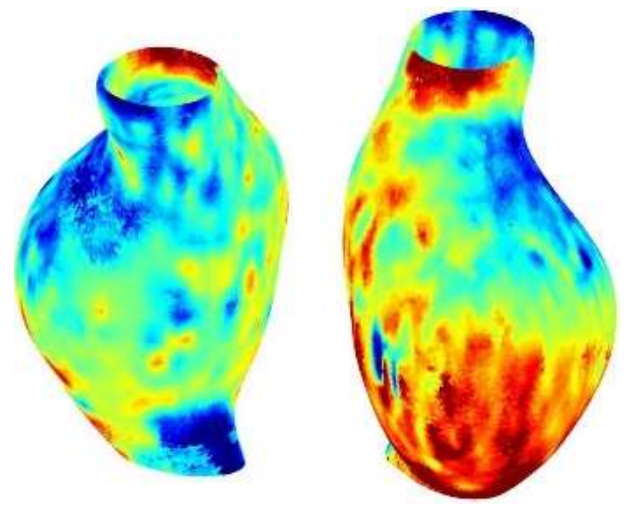<br>$\epsilon_\circ = 3.73\ \%$ | 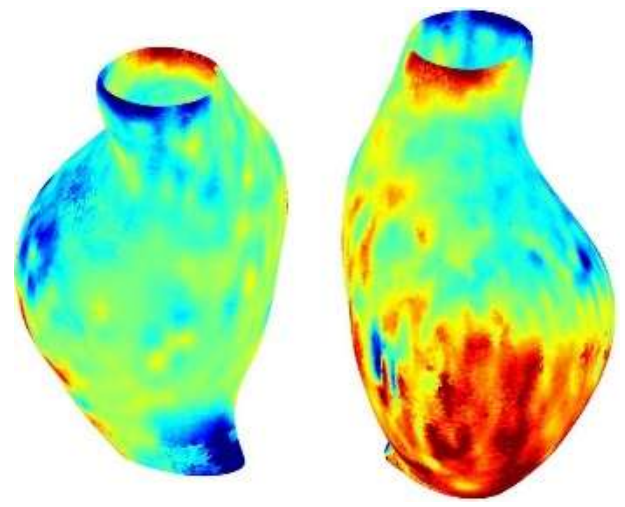<br>$\mathrm{SII}_\circ = 0.35\ \mathrm{mm/N}$ | 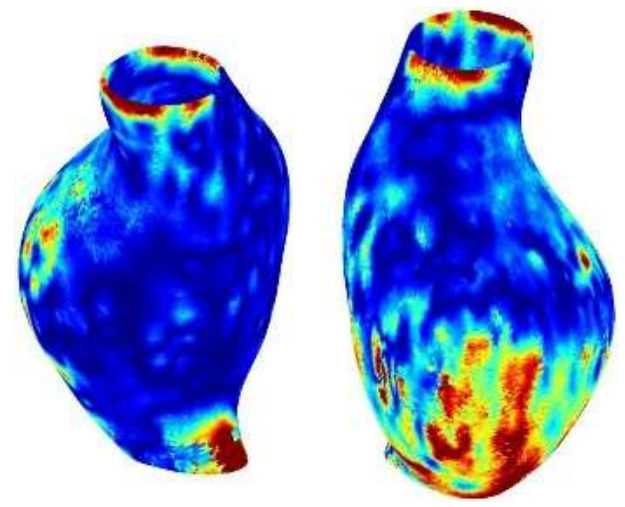<br>$\mathrm{RSII}_\circ = 3.37$ |
| 14 | 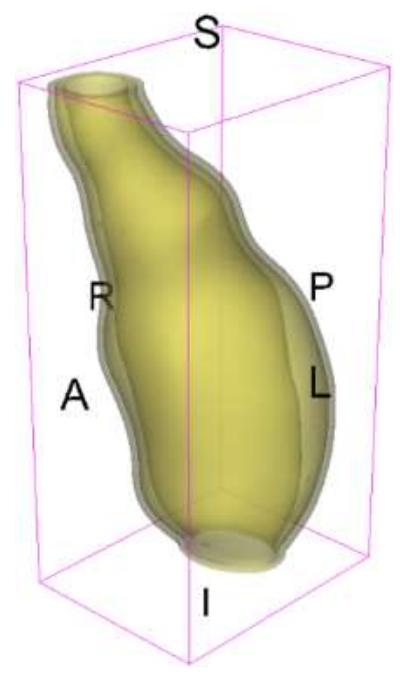 | 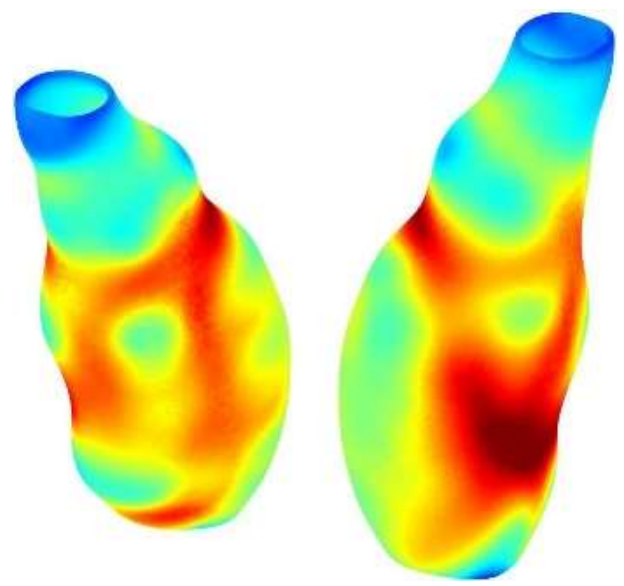<br>$t_\circ = 0.38\ \mathrm{N/mm}$ | 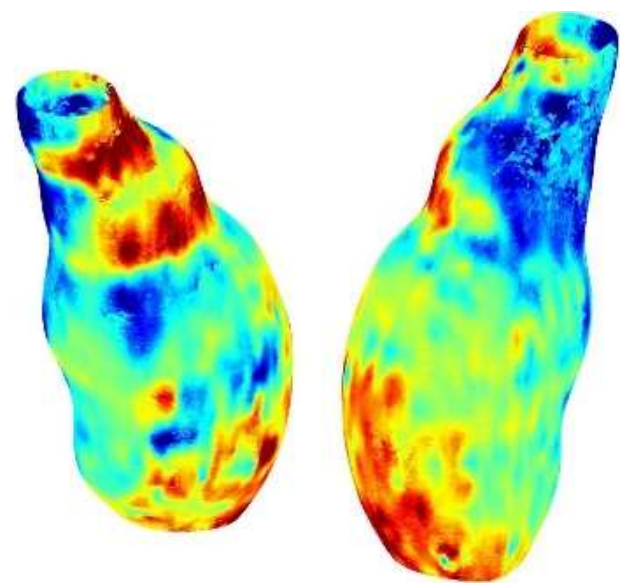<br>$\epsilon_\circ = 5.81\ \%$ | 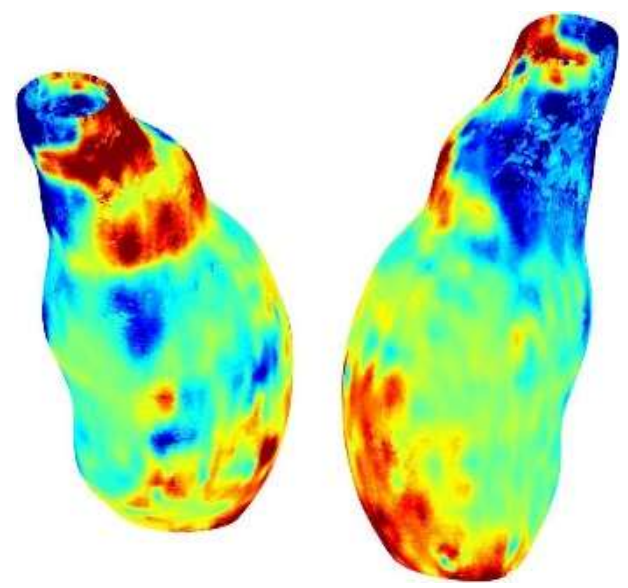<br>$\mathrm{SII}_\circ = 0.29\ \mathrm{mm/N}$ | 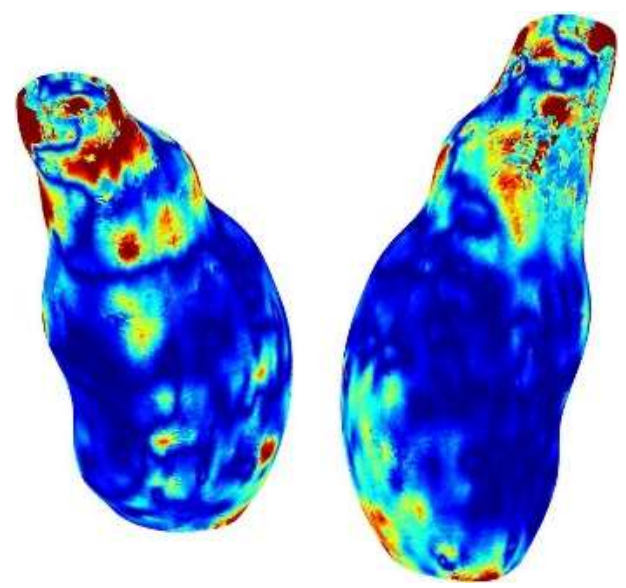<br>$\mathrm{RSII}_\circ = 3.66$ |
| 15 | 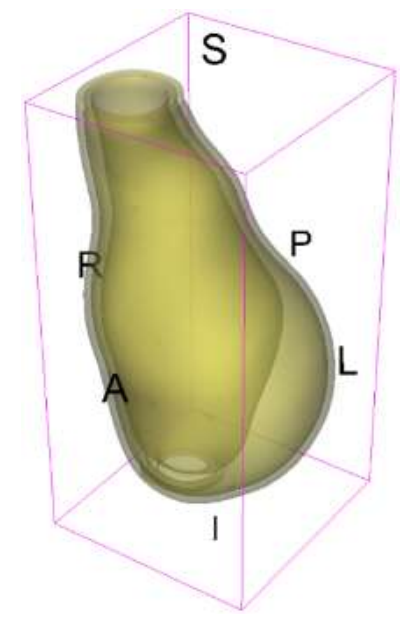 | 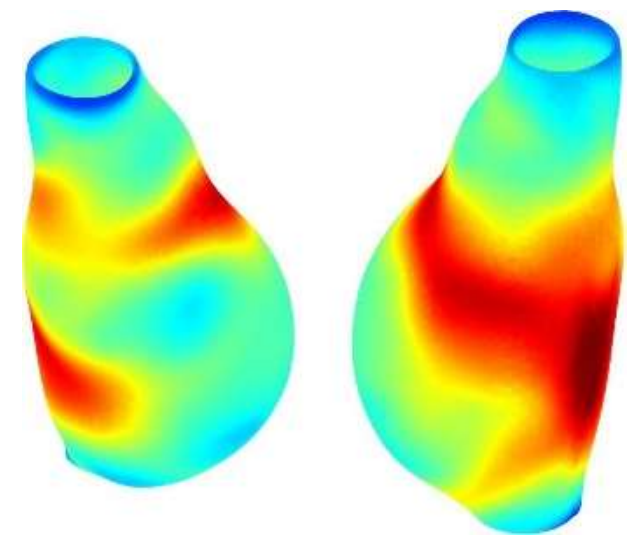 $t_\circ = 0.35$ N/mm | 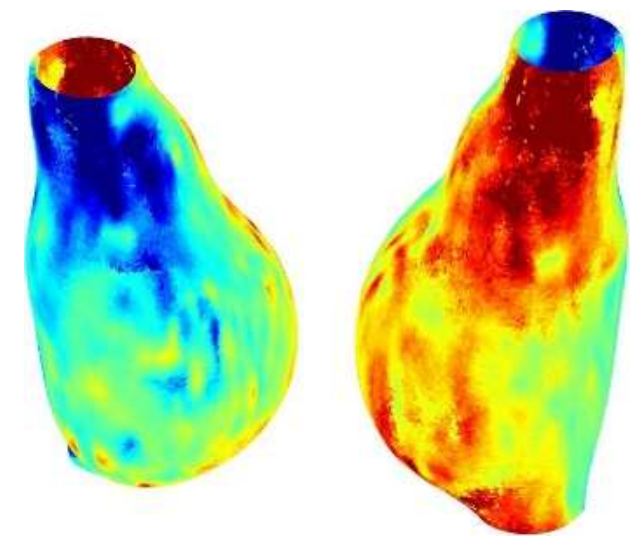 $\epsilon_\circ = 5.43$ % | 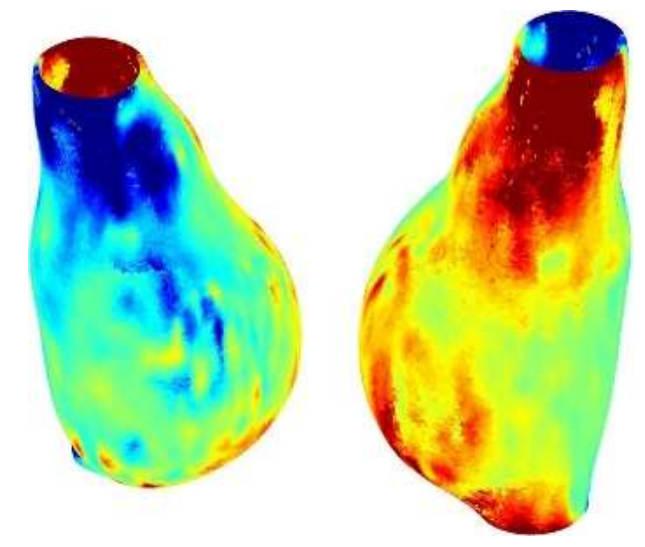 $\mathrm{SII}_\circ = 0.30$ mm/N | 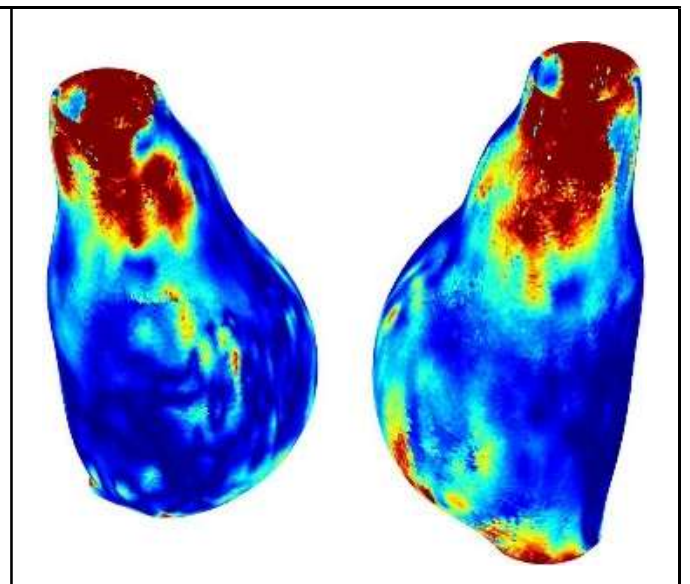 $\mathrm{RSII}_\circ = 3.08$ |
| 16 | 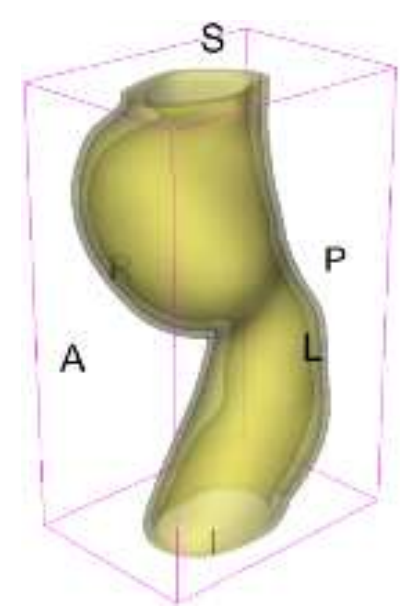 | 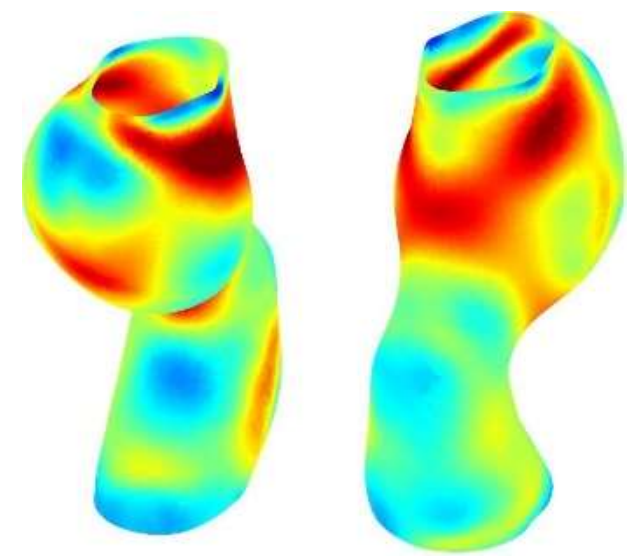 $t_\circ = 0.36$ N/mm | 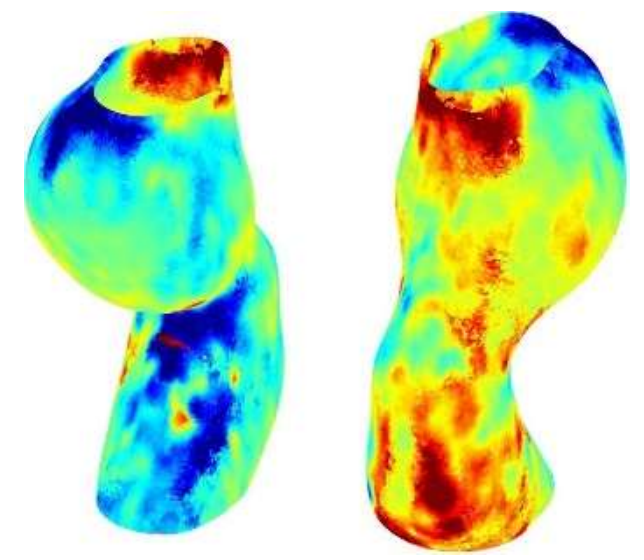 $\epsilon_\circ = 5.43$ % | 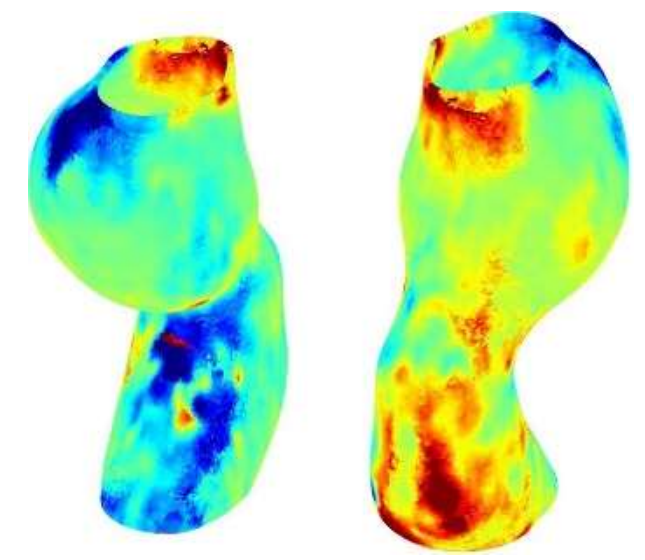 $\mathrm{SII}_\circ = 0.40$ mm/N | 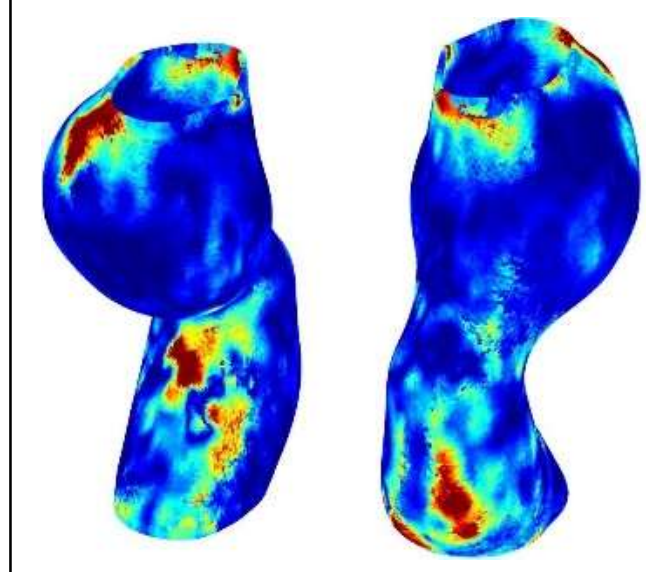 $\mathrm{RSII}_\circ = 4.88$ |
| 17 | 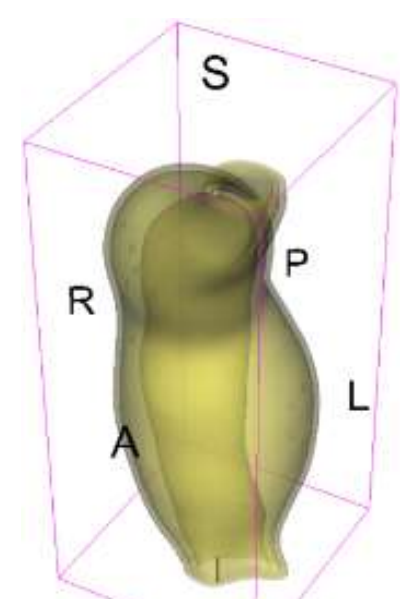 | 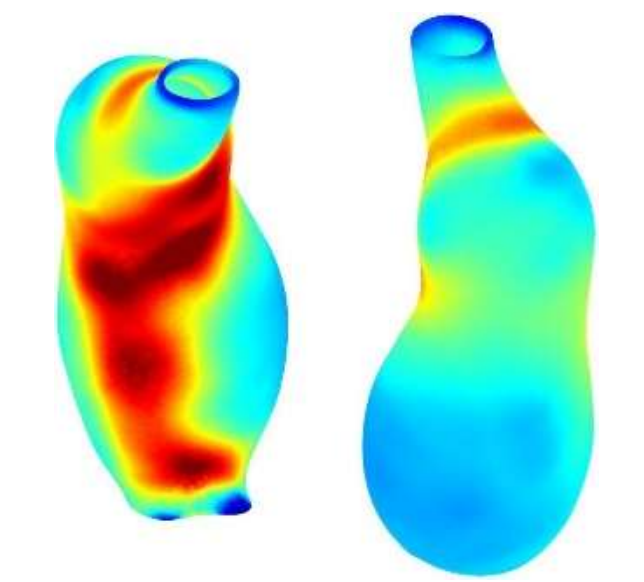 $t_\circ = 0.30$ N/mm | 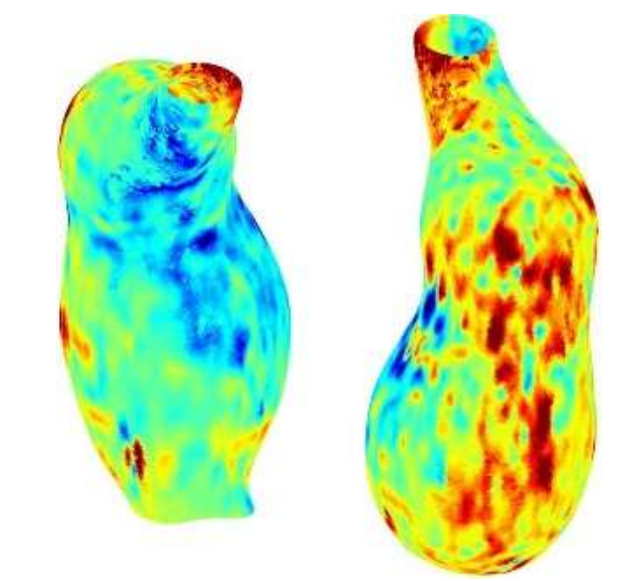 $\epsilon_\circ = 6.08$ % | 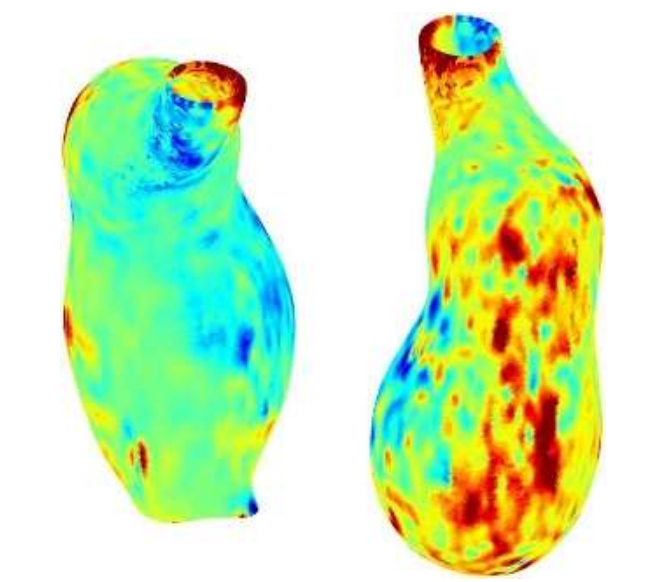 $\mathrm{SII}_\circ = 0.55$ mm/N | 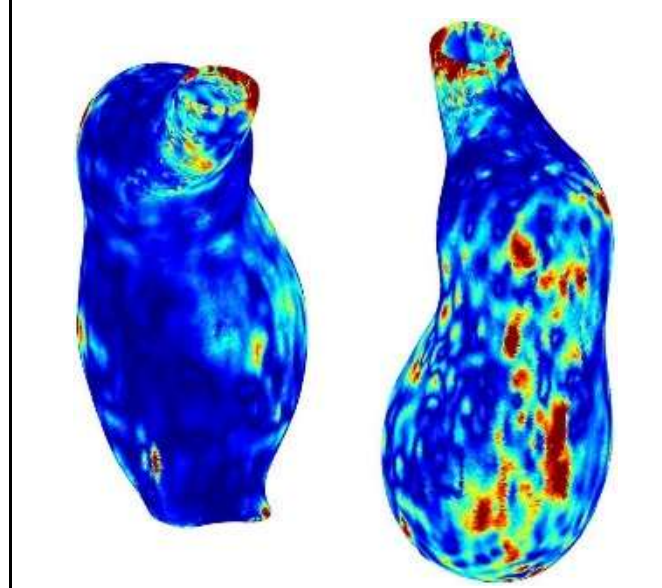 $\mathrm{RSII}_\circ = 4.38$ |
| 18 | 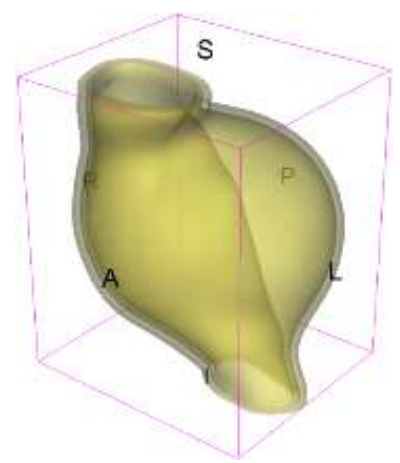 | 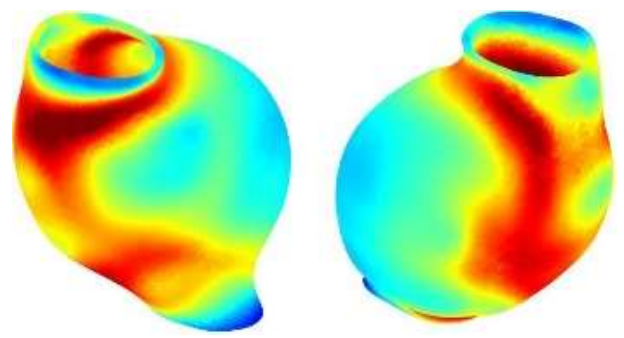 $t_\circ = 0.28$ N/mm | 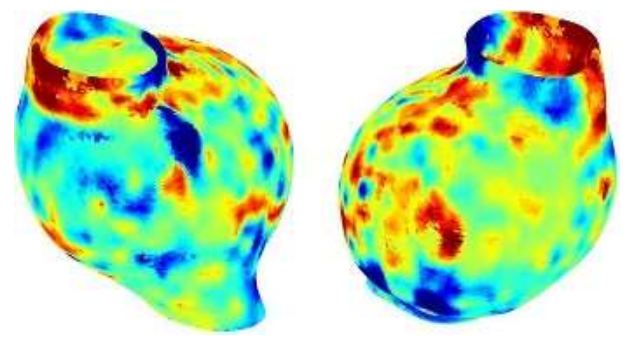 $\epsilon_\circ = 4.79$ % | 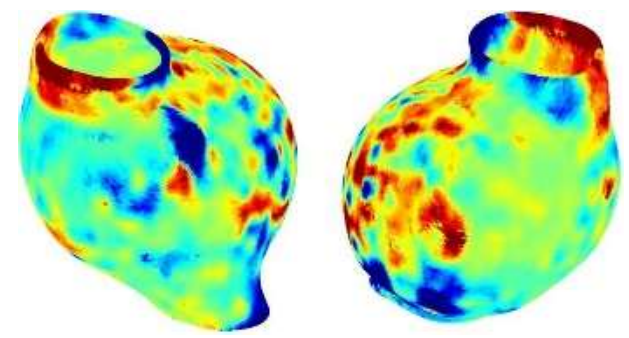 $\text{SII}_\circ = 0.34$ mm/N | 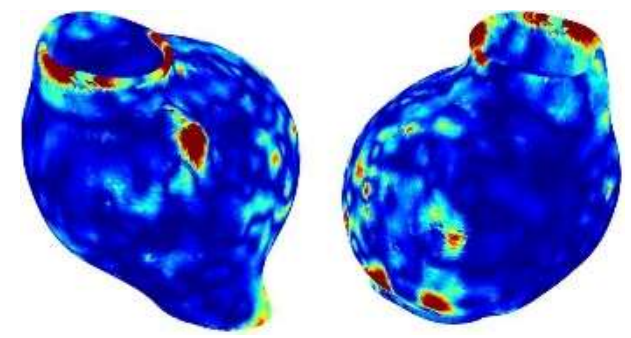 $\text{RSII}_\circ = 5.96$ |
| 19 | 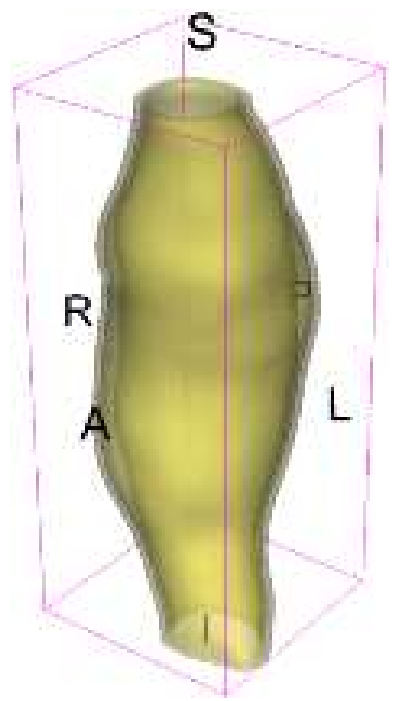 | 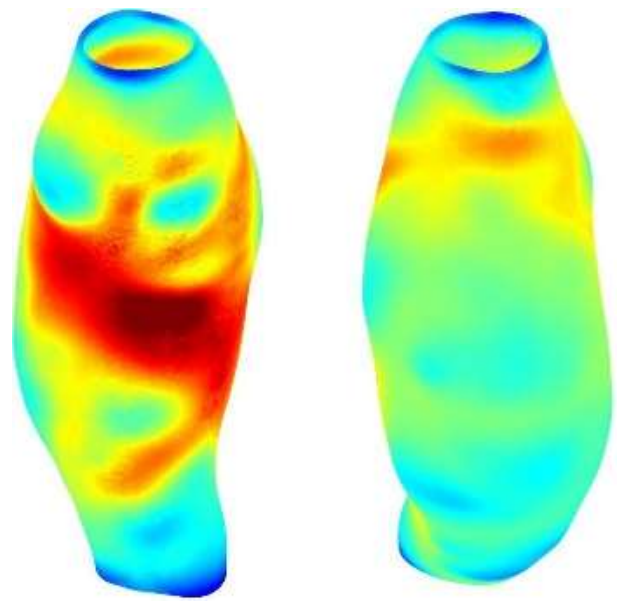 $t_\circ = 0.44$ N/mm | 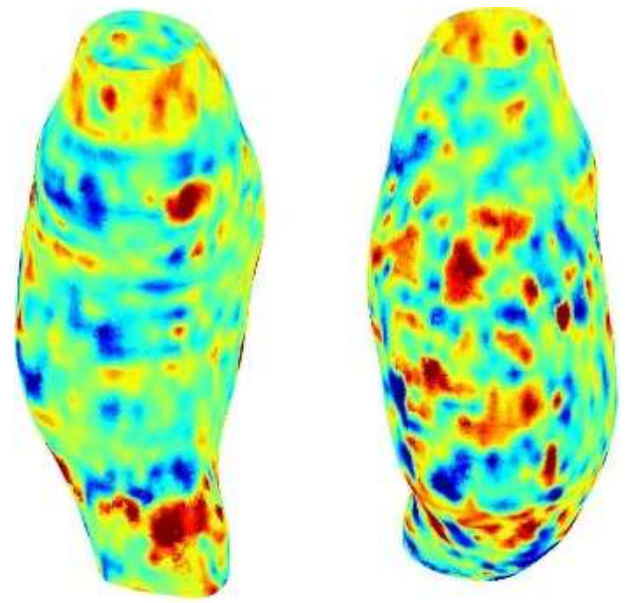 $\epsilon_\circ = 8.32$ % | 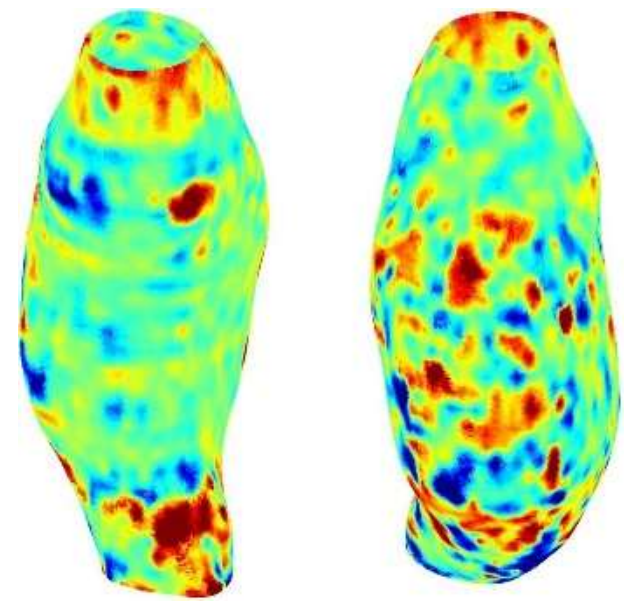 $\text{SII}_\circ = 0.39$ mm/N | 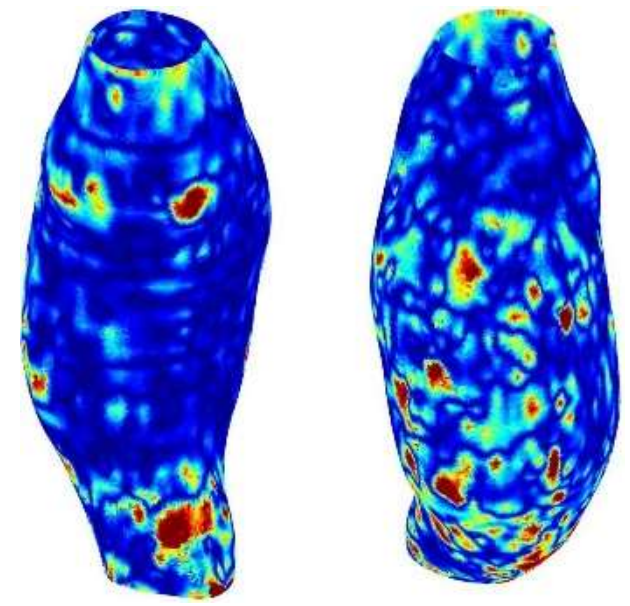 $\text{RSII}_\circ = 4.62$ |
| 20 | 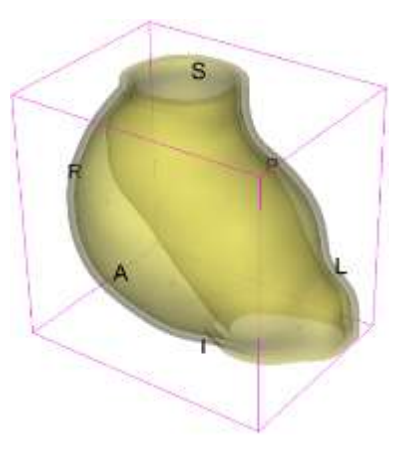 | 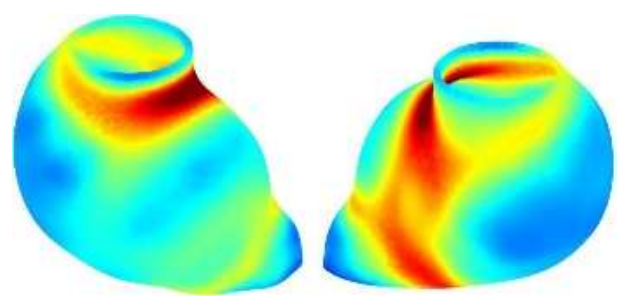 $t_\circ = 0.36$ N/mm | 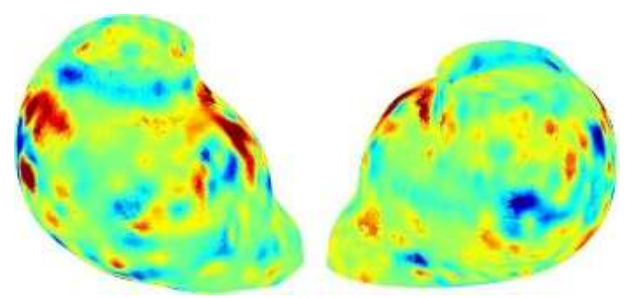 $\epsilon_\circ = 5.11$ % | 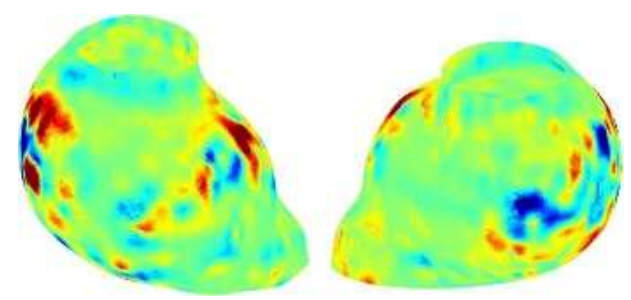 $\text{SII}_\circ = 0.40$ mm/N | 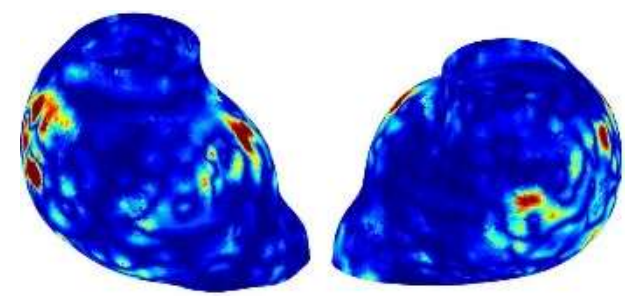 $\text{RSII}_\circ = 6.62$ |

Table 2 shows that the tension and strain maps for each patient do not correlate, as high-stress regions do not align with high-strain regions, and vice versa, suggesting that pure stress-based and pure strain-based rupture risk estimations do not agree. Table 2 shows that, among all patients, the 99th percentile tension varied from 0.25 N/mm to 0.49 N/mm, with an average of 0.35 N/mm and a standard deviation of 0.06 N/mm, while the 99th percentile circumferential strain ranged from 2.88% to 8.32%, with an average of 5.23% and a standard deviation of 1.22%. For a detailed discussion on stress and strain in AAA, readers may refer to our earlier works (Jamshidian et al., 2025b; Miller et al., 2020; Wittek et al., 2022).

The SII maps in Table 2 reveal a non-uniform distribution of wall compliance (i.e. the inverse of stiffness). Localized islands with high absolute SII values may indicate areas of local wall weakening, potentially increasing the likelihood of rupture in these regions. Conversely, regions with low absolute SII values suggest increased wall stiffness (or decreased wall compliance), likely due to local wall strengthening and possibly the presence of calcifications. Positive SII values represent wall compliance calculated based on positive (tensile) strain, while negative SII values indicate wall compliance calculated from negative (compressive) strain.

The RSII maps in Table 2 highlight the high-RSII regions on the AAA surface as potentially rupture-prone locations. In most AAAs, high-RSII regions, surrounded by lower RSII regions, are in the most dilated region of the AAA, which may indicate a higher likelihood of rupture in these regions. In some AAAs, such as those of patients 5, 6 and 8, high-RSII regions are also visible outside the most dilated region of the AAA. This is consistent with experiments on AAAs harvested during autopsy and inflated to rupture, which revealed that one-fourth of the specimens did not rupture in their most dilated region (Gomes et al., 2021). Additionally, more uniform and expanded high-RSII regions are identifiable outside the AAA region, towards the non-aneurysmal healthy aorta. A discussion on RSII in the non-aneurysmal aorta is provided in the next Section.

### 3.2. Healthy aorta structural integrity

Uniaxial tensile testing of AAA wall tissue obtained during autopsy or surgery has established that AAAs are stiffer than non-aneurysmal aortas due to a decreased elastin-to-collagen ratio (He and Roach, 1994). Similarly, recent experiments on AAAs, harvested during autopsy and inflated to rupture, have shown that normal aortas are more compliant than AAAs (Gomes et al., 2021).

To compare RSII in the AAA with that in the healthy portions of the aorta, we performed a structural integrity analysis for Patient 1 using uncropped images that included the healthy proximal sections of the aorta above the AAA region, as shown in Table 3. The contour plots of wall tension and circumferential strain in Table 3 demonstrate that, as discussed in our previous works (Jamshidian et al., 2025b; Miller et al., 2020; Wittek et al., 2022), AAA wall strains are significantly lower than those of a healthy aorta, while higher stresses are developed in the AAA region, compared to the healthy aorta, due to its inflated geometry.

The SII contour plots in Table 3 reveal regions with high absolute SII values in both the aneurysmal aorta and, notably, the non-aneurysmal aorta. The RSII contour plots in Table 3 reveal a more uniform RSII distribution in the healthy aorta proximal to the AAA compared to the AAA itself. Additionally, the non-aneurysmal aorta generally exhibits higher RSII values, reflecting higher compliance (or lower stiffness) compared to the aneurysmal aorta. This is consistent with the experimental findings reported in the literature (Gomes et al., 2021; He and Roach, 1994).

**Table 3.** Structural integrity analysis of healthy aorta for Patient 1. The AAA and healthy aorta geometry, and contour plots of wall tension, circumferential strain, Structural Integrity Index (SII) and Relative Structural Integrity Index (RSII) are presented.

| Healthy aorta and AAA geometry | Wall tension (N/mm) | Circumferential strain | SII (mm/N) | RSII |
|---|---|---|---|---|
| 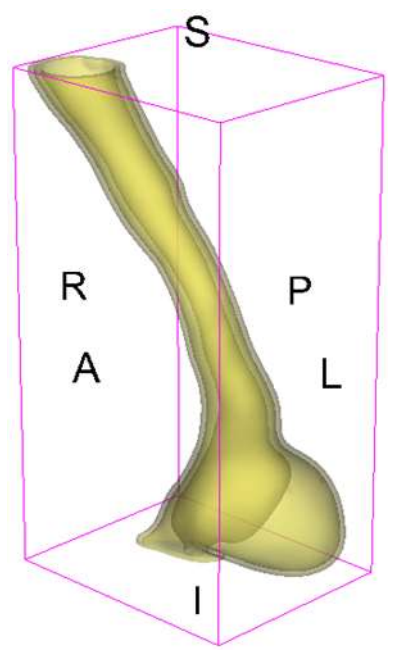 | 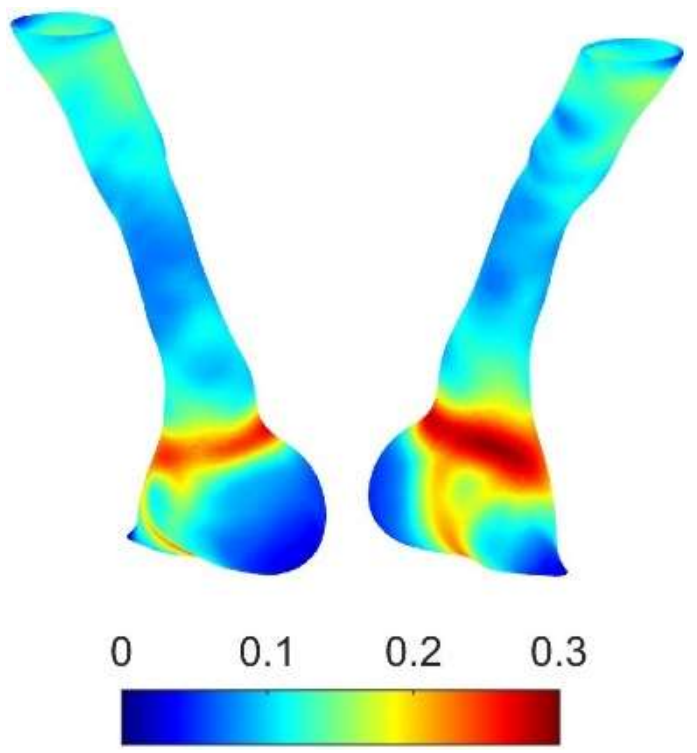 | 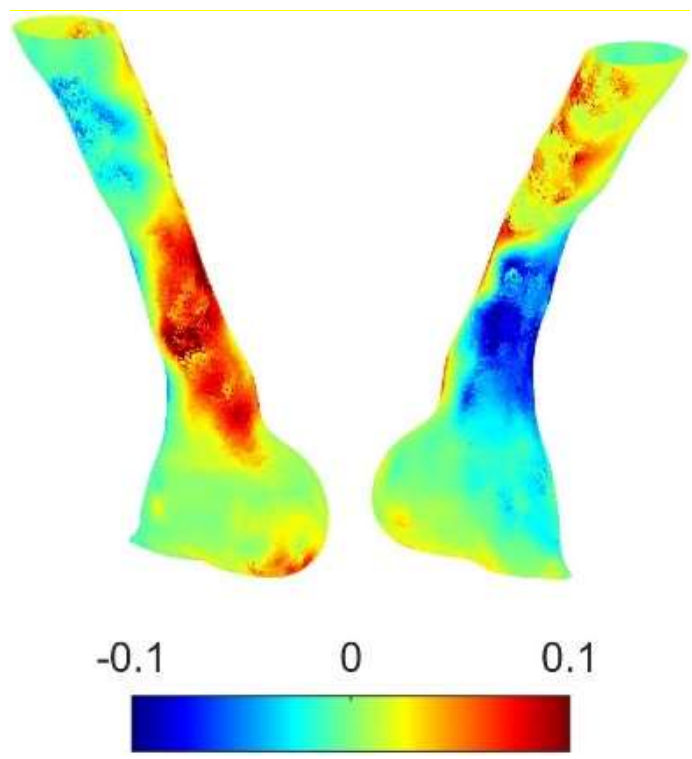 | 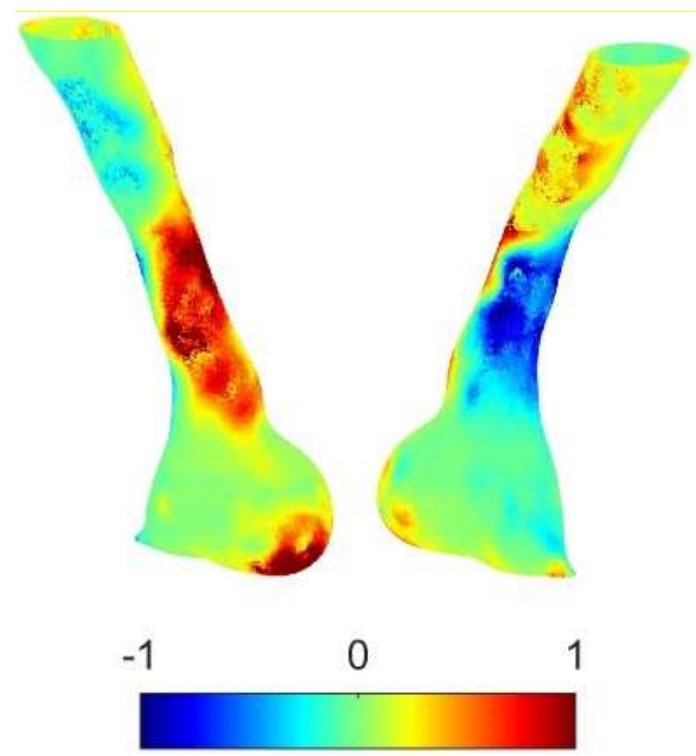 | 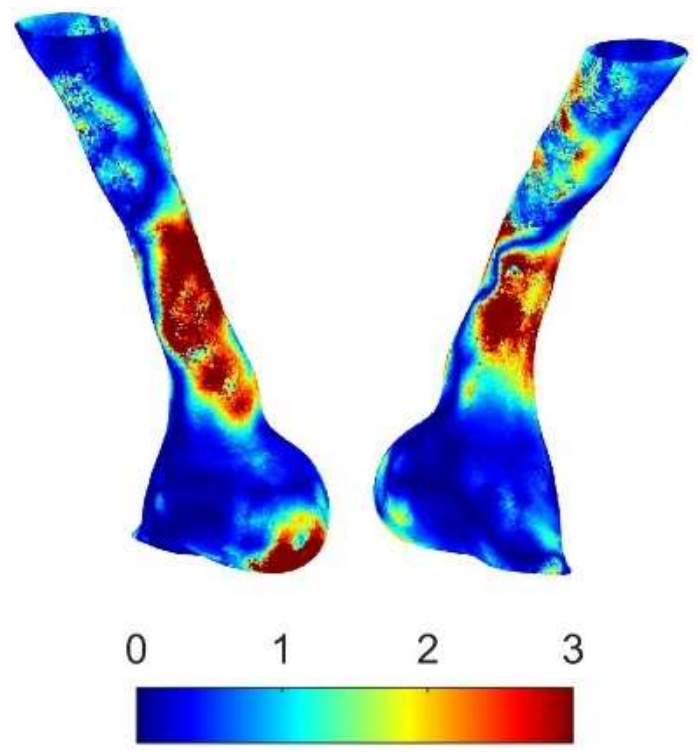 |

## 4. Discussions and conclusions

We proposed a novel measure of AAA wall structural integrity, the Relative Structural Integrity Index (RSII), by integrating computational biomechanics for AAA wall stress recovery with computational image-based kinematics. We calculated the Structural Integrity Index (SII) across the AAA wall as the ratio of circumferential strain to wall tension and defined the RSII as the ratio of the absolute value of SII at a given point to the average absolute value of SII across the AAA wall.

For stress computations using patient-specific image-based FE models, we used the automated pipeline in BioPARR (Biomechanics-Based Prediction of Aneurysm Rupture Risk) software package (Joldes et al., 2017). BioPARR takes the patient's blood pressure along with AAA and blood label maps segmented from medical images, recovers the stress in the AAA wall, and, in the post-processing stage, corrects the recovered stresses to incorporate the effect of residual stresses and computes wall tension. The resulting wall tension does not depend on the assumed properties of the wall tissue, nor the assumed wall thickness used in the calculations.

For strain computations from the patient's 4D-CTA images, we used a MATLAB implementation of deformable image registration with isotropic total variation regularization of displacement (Vishnevskiy et al., 2017). This approach first computes wall displacement from registration and then calculates the circumferential strain as the ratio of normal (to the AAA surface) displacement to the local radius of curvature.

We applied our methods to evaluate wall stress, tension, strain, and RSII in twenty AAAs. We also compared RSII between AAA and the non-aneurysmal aorta. In line with the published experimental observations (Gomes et al., 2021; He and Roach, 1994), our RSII maps revealed the following:

- In most AAAs, localized high-RSII islands (i.e., regions of high strain and low tension) surrounded by lower RSII regions, are identifiable in the most dilated region of the AAA, potentially indicating rupture-prone areas of local wall weakening.
- In some AAAs, high-RSII regions are also visible outside the most dilated region.
- In all AAAs, low-RSII regions in the inflated areas suggest increased wall stiffness, possibly due to the presence of calcifications.

- The more uniform high-RSII distribution in the healthy aorta, compared to the AAA wall, indicates that the healthy aorta is more compliant than the AAA wall.

Given its agreement with experimental findings, RSII has the potential to be used for non-invasive in vivo estimation of AAA disease progression, possibly including rupture risk, in individual patients. RSII addresses the key challenges of patient-specific biomechanical analysis through its two key properties: bypassing the need for patient-specific AAA wall mechanical properties and thickness, as well as precise blood pressure measurement.

RSII is independent of material properties because stress recovery in the known, deformed configuration is a linear problem that is insensitive to the properties of the continuum under consideration (Ciarlet, 2021), as demonstrated for the case of AAA, e.g., in (Joldes et al., 2016).

RSII is independent of the difficult-to-measure AAA wall thickness because, instead of stress, it uses wall tension—a stress resultant calculated from the stress tensor by "integrating the thickness out".

RSII is independent of pressure measurement conditions because it is a relative measure. As a result, any variations in blood pressure and its fluctuations over the cardiac cycle, which may proportionally affect wall tension and SII, are automatically compensated for when computing RSII.

Despite its advantages, our method has certain limitations that should be acknowledged. In current clinical practice, 4D-CTA is not commonly used for AAA diagnosis and treatment. However, given its widespread application in cardiology, it holds strong potential for integration into vascular disease management. Nonetheless, it is worth noting that the proposed methodology, in principle, can be applied to any 4D imaging modality that enables the measurement of displacements between systolic and diastolic geometries, such as 4D-CTA, 4D ultrasound, or CINE MRI.

It should be emphasized that the present study is primarily methodological and does not aim to provide statistically significant validation of RSII as a biomarker. The proposed methodology is ready for implementation but has not yet been clinically validated in terms of demonstrating benefit to patients. To achieve such clinical validation, we plan to conduct a multi-centre, multi-year observational study, with the first results expected in 2027.

Further work is required to properly interpret RSII distribution over the AAA surface and perhaps suggest a single numerical indicator with predictive power. To investigate the utility of RSII in AAA disease progression and rupture risk estimation, future longitudinal cohort studies of AAA patients under surveillance should examine the correlation between RSII-based AAA assessment and clinically observed disease progression.

## Acknowledgements

This work was supported by the Australian National Health and Medical Research Council NHMRC Ideas grant No. APP2001689. Contributions of Christopher Wood and Jane Polce, radiology technicians at the Medical Imaging Department, Fiona Stanley Hospital (Murdoch, Western Australia) to patient image acquisition at Fiona Stanley Hospital, and Dr Farah Alkhatib's (The University of Western Australia) assistance in the acquisition of these images are gratefully acknowledged.